\documentclass[journal]{IEEEtran}
\usepackage{cite}
\usepackage{hyperref}

\ifCLASSINFOpdf
\else
\fi
\usepackage{stackengine}
\usepackage{amsmath}
\usepackage{amsfonts}
\usepackage{siunitx}

\usepackage{array}
\makeatletter
\let\MYcaption\@makecaption
\makeatother

\usepackage{silence}
\usepackage[font=footnotesize]{subcaption}

\makeatletter
\let\@makecaption\MYcaption
\makeatother

\usepackage{multirow}

\usepackage{amsthm}
\usepackage{graphicx}
\graphicspath{ {./images/} }

\usepackage[section]{placeins}
\usepackage{float}

\usepackage{pgfplots}
\pgfplotsset{compat=newest}
\usetikzlibrary{calc}
\usetikzlibrary{plotmarks}
\usetikzlibrary{arrows.meta}
\usetikzlibrary{external}
\usetikzlibrary{decorations.shapes} 
\usepgfplotslibrary{patchplots}
\usepackage{grffile}
\usepackage{amsmath}
\makeatletter
\long\def\ifnodedefined#1#2#3{%
    \@ifundefined{pgf@sh@ns@#1}{#3}{#2}%
}
\makeatother

\usepackage{tikzscale}

\usetikzlibrary{positioning}

\usepackage{xcolor}

\definecolor{myorange}{rgb}{0, 0.5, 0}
\definecolor{myturkis}{rgb}{0.8, 0, 1}
\definecolor{myblue}{rgb}{0, 0, 1}

\begin{document}

\DeclareRobustCommand{\lineleg}[1]{\tikz[baseline=1.7ex]\draw[#1, line width=0.4mm] (0,0.3) -- (0.5,0.3);} 

\newcommand{\annotateChangesStart}[0]{}
\newcommand{\annotateChangesEnd}[0]{}

\twocolumn[
\begin{@twocolumnfalse}

\setcounter{page}{0}

\textbf{Author's post-print} \\

%This work has been submitted to the IEEE for possible publication. Copyright may be transferred without notice, after which this version may no longer be accessible.

\copyright 2024 IEEE. Personal use of this material is permitted. Permission from IEEE must be obtained for all other users, in any current or future media, including reprinting/republishing this material for advertising or promotional purposes, creating new collective works, for resale or redistribution to servers or lists, or reuse of any copyrighted components of this work in other works.\\\\
DOI: \url{https://doi.org/10.1109/TAP.2024.3515285} \\ %fill this for post-print
%URL: \url{}\\ %fill this for post-print

%Cite: L. Mörlein, D. Manteuffel, “Array  Synthesis  in Terms of Characteristic Modes and Generalized Scattering Matrices,” \textit{IEEE Transactions on Antennas and Propagation}, vol. xx, no. xx, id. xx, Jan. 2021. \\ %fill this for post-print

\end{@twocolumnfalse}
]

%\title{Modeling of Mutual Coupling in Antenna Arrays\\Using Generalized Scattering Matrices \\in Terms of Characteristic Modes}
%\title{Modeling of Mutual Coupling in Antenna Arrays\\Using GSM in Terms of Characteristic Modes}
%\title{A Model for Mutual Couping in Antenna Arrays\\ in Terms of Characteristic Modes}
\title{Array  Synthesis  in Terms of Characteristic Modes and Generalized Scattering Matrices}
%\title{Enhancing the Active Modal Configuration of a Circular Polarized Patch Antenna Array Using Characteristic Modes} % ICEAA Paper

\author{Leonardo~M\"orlein,~\IEEEmembership{Student Member,~IEEE,}
        and~Dirk~Manteuffel,~\IEEEmembership{Member,~IEEE}% <-this % stops a space
\thanks{The authors are with the Institute of Microwave and Wireless Systems, Leibniz University Hannover, Appelstr. 9A, 30167 Hannover, Germany (e-mail:
moerlein@imw.uni-hannover.de; manteuffel@imw.uni-hannover.de).}% <-this % stops a space
}


%\markboth{IEEE TRANSACTIONS ON ANTENNAS AND PROPAGATION,~Vol.~X, No.~Y, MM~2024}%
%{Mörlein \MakeLowercase{\textit{et al.}}: Array Coupling in Terms of Characteristic Modes and Generalized Scattering Matrices}
\markboth{}%
{Mörlein \MakeLowercase{\textit{et al.}}: Array Synthesis in Terms of Characteristic Modes and Generalized Scattering Matrices}



\maketitle

\begin{abstract}

The synthesis of antenna arrays in presence of mutual coupling using generalized scattering matrices in terms of characteristic modes is proposed.
For the synthesis, the array is built of synthetic elements that are described by their modal scattering and radiation behavior.
In particular, the question of how to describe the degrees of freedom of such elements is addressed.
The eigenvalues of the characteristic modes of the element geometry and the modal radiation behavior of the antenna are thereby selected as degrees of freedom for the model of the synthetic elements.
Using this model and a modal coupling matrix, an approach to optimize the modal configuration of the elements within an array is proposed.
Finally, a close to reality example shows how the proposed theory can be used to enhance the cross-polarization rejection of a circularly polarized patch antenna array with a fixed beam.
\end{abstract}


% \fi
\IEEEpeerreviewmaketitle

\section{Introduction}
%\IEEEPARstart{T}{his} demo file is intended to serve as a ``starter file''
%for IEEE journal papers produced under \LaTeX\ using
%IEEEtran.cls version 1.8b and later.
%I wish you the best of success.

% Keinen größeren Zeilenabstand erzwingen, um die Seite zu füllen.
\raggedbottom 

The performance of an antenna array is highly influenced by the mutual coupling of the individual antenna elements in that antenna array. Therefore, modeling of the mutual coupling is an important aspect in the design of antenna arrays \cite{Bird2021}.

To evaluate the mutual coupling between antenna elements, a full-wave simulation of the entire array is often performed. On the one hand, a full-wave simulation gives accurate results, but on the other hand, it delivers no insight on how different design parameters influence the mutual coupling. Since the design of antenna arrays requires control of the mutual coupling, antenna designers are looking for  mathematical  models to describe the {}mutual coupling between antenna elements in terms of underlying parameters that can be used for design.

Often, these models rely on the use of port-based quantities. For example, the optimization of the port excitation \cite{Capek2021a} or the optimization of termination impedances for parasitic scatterers \cite{Salmi2024,Salmi2024a} can be solved using port-based quantities. However, sometimes the port position or even the shape of the antenna itself is subject of optimization. For these situations, it is useful to consider a description in the form of higher order quantities, such as modes.

In that direction, e.g. Wasylkiwskyj and Khan have addressed the question how mutual coupling in antenna arrays can be calculated based on prescribed radiation patterns in terms of spherical wave functions \cite{Wasylkiwskyj1970,Wasylkiwskyj1970a,Rubio2011,Ghosal2021}.
To achieve a result which is solely dependent on the prescribed radiation patterns, they restrict their definition to minimum scattering antennas~\cite{Kahn1965}.

Although this approach has existed for a long time now, mutual coupling is still mainly evaluated using full-wave simulations today.
One reason for this could be that spherical wave functions are not specific to a particular element geometry, but result from a general decomposition of the Helmholtz equation.
This complicates the formulation of the degrees of freedom for a specific element geometry using spherical wave functions, thereby making it challenging to derive conclusions about the geometric parameters of the antenna elements.

An upcoming approach to obtain more insight into the coupling phenomena is to break down the coupling between ports into the coupling between the so-called characteristic modes \cite{Lou2019,Ghosal2020}.
The characteristic modes provide an eigensolution to the electromagnetic scattering fields of a particular element geometry \cite{Harrington1971} and therefore form a good basis to describe radiation and scattering problems \cite{Adams2022}.

In this work, a mathematical approach to the synthesis of antenna arrays with synthetic array elements is presented. As model for the synthetic antenna elements, a generalized scattering matrix in terms of the modal scattering and radiation parameters of the antenna is selected. By using characteristic modes as basis for the synthesis, a compact description of the degrees of freedom is found. This description even allows the consideration of antennas of a more general antenna type than minimum scattering antennas, as described for example by Wasylkiwskyj and Khan.

%{ \color{red}
%This model is inspired by the above-mentioned work of Wasylkiwskyj and Khan. However, firstly, we use characteristic modes as modal decomposition to obtain a sparser representation of the modal antenna properties. Secondly, we extend the scope of the synthesis to a more general type of antennas than minimum scattering antennas.
%}

%On the one hand, the proposed approach can be used as a basis for mutual coupling calculation methods. On the other hand, the parameters introduced by the model provide a new intermediate layer to understand coupling phenomena at a higher and more abstract level.
%For example, it becomes possible to answer questions like 'How much would ports on adjacent elements couple if I were to realize ports that excite only mode 2 on each element?' for a particular array without actually realizing a port that is impedance matched and excites only mode 2. This could be particularly interesting for example for the design of innovative multi-mode multi-port antenna arrays, where each array element typically has four or more ports that excite different modes on each element \cite{Manteuffel2016,Moerlein2021,Peitzmeier2022,Manteuffel2022}.

The structure of the paper is as follows: In section~\ref{sec:theory_isolated}, a method for a mathematical synthesis of isolated antenna elements in terms of generalized scattering matrices is introduced. In section~\ref{sec:theory_array_level}, a formalism to describe mutual coupling in arrays in terms of these generalized scattering matrices is introduced. In section~\ref{sec:example}, the proposed theory is applied to describe circularly polarized patch elements within an array and to optimize the performance of that array under the consideration of mutual coupling. Section~\ref{sec:conclusion} concludes with a summary of the work and an outlook.

\section{Considerations on the Isolated Element Level}
\label{sec:theory_isolated}

In order to optimize the coupling between the elements of the antenna array, a mathematical model of the elements of the array is first developed in this section. The elements are thereby considered as isolated from the other elements. Therefore, the eigenmodes of a single isolated antenna are first introduced in the following and linked to a generalized scattering matrix, which enables the subsequent introduction of mutual coupling in the next section.

\subsection{Characteristic Modes}

%\subsection{Definitions}

%\subsubsection{The Eigenproblem of Characteristic Modes}

%For spherical wave functions, the calculation of the scattering matrix $\mathbf{S}_0$ requires the analytical formula of spherical wave functions to be implemented in the method of moments code and can be calculated as described in \cite{Gustafsson2022a}.

%For characteristic modes, the calculation does not require an additional analytical formula.

If an electric field integral equation scheme of the method of moments is considered, the characteristic modes are defined based on a generalized eigenvalue problem of the impedance matrix $\mathbf{Z}_0$\footnote{\annotateChangesStart The impedance matrix $\mathbf{Z}_0$ is the method of moments impedance matrix without port termination loads. \annotateChangesEnd}:
\begin{equation}
\label{eq:z0_cm_definition}
    \operatorname{Im}\left\{\mathbf{Z}_{0}\right\} \mathbf{I}_{\mathrm{CM}} = \operatorname{Re}\left\{\mathbf{Z}_{0}\right\} \mathbf{I}_{\mathrm{CM}} \mathbf{\Lambda}_0.
\end{equation}
Hereby $\mathbf{\Lambda}_0$ is a diagonal matrix with the eigenvalues $\lambda_n$ on the diagonal and $\mathbf{I}_{\mathrm{CM}}$ is a matrix that contains the characteristic mode current distributions $\mathbf{J}_n$ represented in the basis functions of the method of moments as column vectors. Throughout this paper, the currents are thereby normalized such that $\mathbf{I}_{\mathrm{CM}}^\mathrm{T} \operatorname{Re}\left\{\mathbf{Z}_{0}\right\} \mathbf{I}_{\mathrm{CM}} = \mathbf{I}$, whereby $\mathbf{I}$ denotes a unit matrix.

In this formulation, characteristic modes are used to describe a scattering problem. However, when characteristic modes are used to describe elements within an array, the elements act simultaneously as radiators and scatterers \cite{Hansen1988}. Therefore, it makes sense to describe the antenna in terms of a generalized scattering matrix where both is possible simultaneously.

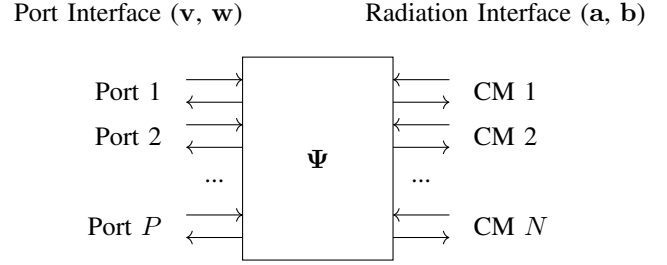
\begin{figure}
\centering
\begin{tikzpicture}
    \pgfmathsetmacro{\boxWidth}{2}
    \pgfmathsetmacro{\arrowCount}{3}
    \pgfmathsetmacro{\arrowDist}{0.3}
    \pgfmathsetmacro{\arrowLength}{0.75}
    \pgfmathsetmacro{\boxHeight}{(3+\arrowCount*2)*\arrowDist}
    \draw[] (0,0) -- (\boxWidth,0) -- (\boxWidth,-\boxHeight) -- (0,-\boxHeight) -- cycle;
    \node at (\boxWidth/2,-\boxHeight/2) {$\mathbf{\Psi}$};
    
    \def\setelem#1{\expandafter\def\csname myarray(#1)\endcsname}
    \def\lookup#1{\csname myarray(#1)\endcsname}
    
    \newcommand{\makeArrows}[1]{
        \foreach \x in {0,1,3} {
            \draw[\arrowDir] (-\arrowLength, -\arrowDist-2*\x*\arrowDist) -- (0, -\arrowDist-2*\x*\arrowDist);
            \draw[\arrowDir] (0, -2*\arrowDist-2*\x*\arrowDist) -- (-\arrowLength, -2*\arrowDist-2*\x*\arrowDist);
            
            \node [anchor=#1] (label\x) at (-\arrowLength+\labelOffset, -1.5*\arrowDist-2*\x*\arrowDist) {\lookup{\x}};
        }
        \pgfmathsetmacro{\x}{2}
        \node at (-\arrowLength/2, -1.5*\arrowDist-2*\x*\arrowDist) {...};
    }

    \def\arrowDir{->}
    \setelem{0}{Port 1}
    \setelem{1}{Port 2}
    \setelem{3}{Port $P$}
    
    \pgfmathsetmacro{\labelOffset}{-0.2}
    \makeArrows{east}
    
    \node[above=5mm of label0] {Port Interface ($\mathbf{v}$, $\mathbf{w}$)};
    
    \def\arrowDir{<-}
    \setelem{0}{CM 1}
    \setelem{1}{CM 2}
    \setelem{3}{CM $N$}
    
    \pgfmathsetmacro{\labelOffset}{\arrowLength+0.2}
    \begin{scope}[shift={(\boxWidth+\arrowLength,0)}]
        \makeArrows{west}
    \end{scope}
    
    \node[above=5mm of label0] {Radiation Interface ($\mathbf{a}$, $\mathbf{b}$)};
    
\end{tikzpicture}
\caption{Illustration of the generalized scattering matrix for an antenna element with $P$ ports that interacts with characteristic modes (CM) up to number $N$.}
\label{fig:gsm_illustation}
\end{figure}

A generalized scattering matrix \cite{Hansen1988} describes the relationship between the coefficients of the incident and reflected waves at the ports and the coefficients of incident and radiating modes of the antennas:
\begin{equation}
\label{eq:gsm_generic}
    \underbrace {\begin{bmatrix}
  {\mathbf{S}}&{\mathbf{T}} \\ 
  {\mathbf{R}}&{\mathbf{\Gamma }} 
\end{bmatrix}}_{\mathbf{\Psi}}\begin{bmatrix}
  {\mathbf{a}} \\ 
  {\mathbf{v}} 
\end{bmatrix} = \begin{bmatrix}
  {\mathbf{b}} \\ 
  {\mathbf{w}} 
\end{bmatrix},
\end{equation}
whereby $\mathbf{a}$ and $\mathbf{b}$ denote the coefficient vectors of the incident and radiating modes at the radiation interface while $\mathbf{v}$ and $\mathbf{w}$ denote the incident and reflected waves at the antenna ports as depicted in Fig.~\ref{fig:gsm_illustation}. The matrix $\mathbf{S}$ is called antenna scattering matrix, the matrix $\mathbf{T}$ is called antenna transmit matrix, the matrix $\mathbf{R}$ is called antenna receive matrix and $\mathbf{\Gamma}$ contains the port scattering parameters of the antenna.

\annotateChangesStart

The antenna transmit matrix $\mathbf{T}$ is composed by the entries $t_{n,p}$ which represent how much the $n$\nobreakdash-th characteristic mode is excited by the $p$\nobreakdash-th port. They can be calculated from the current distribution $\mathbf{I}_p$ according to
\begin{equation}
    t_{n,p} = \frac{1}{1+\mathrm{j}\lambda_n} \mathbf{I}_{\mathrm{CM},n}^{\mathrm{T}} \mathbf{Z}_0 \mathbf{I}_p,
\end{equation}
whereby $\mathbf{I}_p$ is the current distribution when only the $p$\nobreakdash-th port is excited by $v_p = 1$ and the other ports are terminated by their characteristic impedances.

\annotateChangesEnd

It is noted that characteristic modes have already been used in the past to calculate the generalized scattering matrix of an antenna in the basis of spherical wave functions \cite{Kim2013,Kim2018}. However, to obtain a sparse and compact representation, we propose to formulate the generalized scattering matrix directly in the basis of the characteristic mode eigenfields $\mathbf{E}_n$. This means that $\mathbf{a}$ and $\mathbf{b}$ describe the coefficients of incoming and outgoing characteristic mode fields:
\begin{equation}
    \mathbf{E} = \sum_n \mathbf{E}_n b_n + \mathbf{E}_n^* a_n.
\end{equation}

However, while this formulation is now based on the eigenfields of the characteristic modes  $\mathbf{E}_n$, the eigenvalues $\lambda_n$ do not appear directly within the generalized scattering matrix. In order to understand the influence of them, a relationship between the eigenvalues $\lambda_n$ of the element geometry without ports and the generalized scattering matrix of the antenna is found in the following.

\subsection{A Relationship Between an Antenna and the Associated Eigenproblem}

To connect the eigenvalues of the characteristic modes $\lambda_n$ to the generalized scattering matrix of the antenna $\mathbf{\Psi}$, the scattering matrix formulation of the characteristic mode eigenvalue problem from \cite{Harrington1971,Gustafsson2022a} can be used.

In this formulation, the scattering behavior of a scattering object (without ports) is also described in terms of incoming and outgoing characteristic mode fields. The relationship between the coefficients $\mathbf{a}$ and $\mathbf{b}$ is thereby described by a scattering matrix $\mathbf{S}_0$:
\begin{equation}
    \mathbf{b} = \mathbf{S}_0 \, \mathbf{a},
\end{equation}
whereby, the scattering matrix $\mathbf{S}_0$ is defined by the eigenvalues $\lambda_n$ of the scattering object \cite{Harrington1971}:
\begin{equation}
\label{eq:scattering_matrix_definition_cm}
    {{\mathbf{S}}}_0 = \operatorname{diag} \{ s_n \} = \begin{bmatrix}
  { - \frac{{1 - {\text{j}}{\lambda _1}}}{{1 + {\text{j}}{\lambda _1}}}}&0&0&{...} \\ 
  0&{ - \frac{{1 - {\text{j}}{\lambda _2}}}{{1 + {\text{j}}{\lambda _2}}}}&0&{...} \\ 
  {...}&{...}&{...}&{...}
\end{bmatrix}.
\end{equation}

Now, in order to connect $\mathbf{S}_0$ and $\mathbf{\Psi}$, it is assumed in the following that the scattering object (represented by $\mathbf{S}_0$) is the antenna from before (represented by $\mathbf{\Psi}$), whereby all ports have been removed. Technically, this removal of ports is a lossless termination of the antenna, represented by a reflection coefficient matrix $\mathbf{\Gamma}_\mathrm{L,0}$. For example, for dipole-like structures, the characteristic modes are typically analyzed for short-circuited ports (${\mathbf{\Gamma}_{\mathrm{L},0}=-\mathbf{I}}$), whereby for patch antennas, the modes are typically analyzed for open-circuited ports (${\mathbf{\Gamma}_{\mathrm{L},0}=\mathbf{I}}$) according to \cite{Yang2016}.

Inserting $\mathbf{\Gamma}_{\mathrm{L},0}$ into \eqref{eq:gsm_generic} leads to the following system of equations:
\begin{equation}
    \begin{bmatrix}
  {\mathbf{S}}&{\mathbf{T}} \\ 
  {\mathbf{R}}& \mathbf{\Gamma} 
\end{bmatrix} \begin{bmatrix}
\mathbf{a} \\
\mathbf{v}
\end{bmatrix} = \begin{bmatrix}
\mathbf{b} \\
\mathbf{\Gamma}_\mathrm{L,0} \mathbf{v}
\end{bmatrix}.
\end{equation}

\begin{figure*}
    \centering
    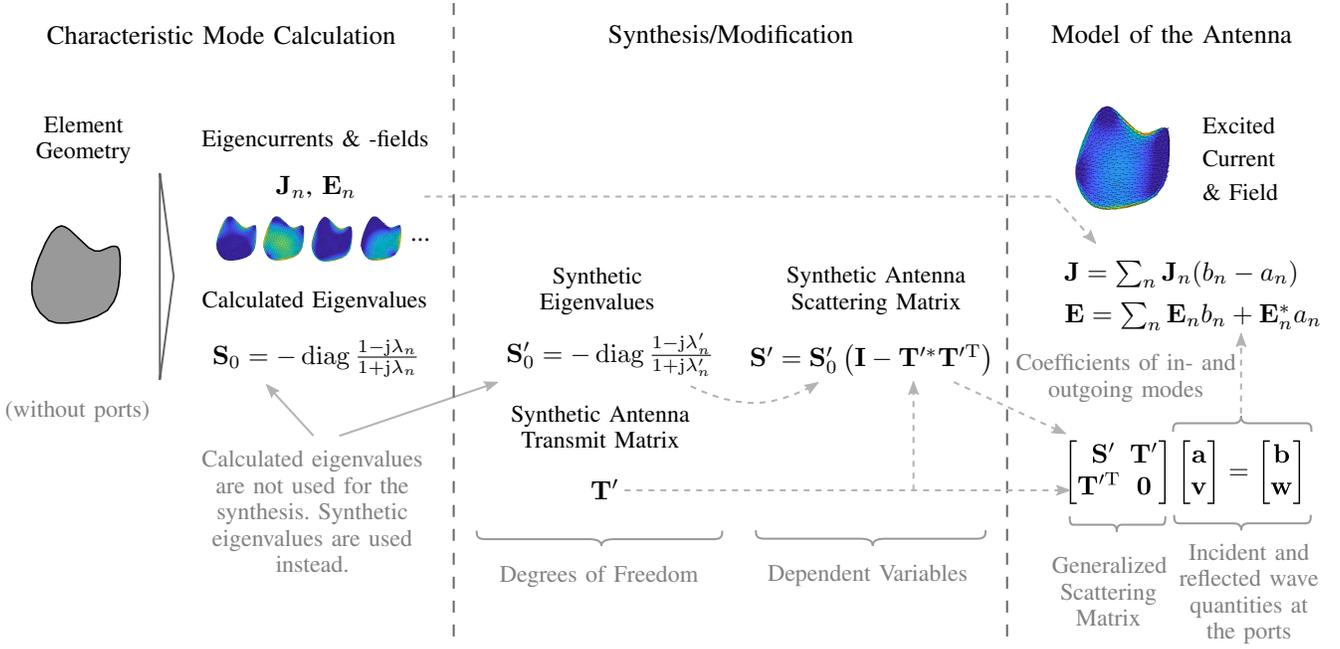
    \caption{ Overview of the proposed antenna synthesis approach based on a given element geometry and synthetic modal antenna parameters that are used as degrees of freedom for the synthesis.}
    \label{fig:overview_synthesis}
\end{figure*}

By combining the upper and lower block of the system of equations, the formula 
\begin{equation}
    \mathbf{b} = \underbrace{\left( \mathbf{S} + \mathbf{T} \left(\mathbf{\Gamma}_\mathrm{L,0}-\mathbf{\Gamma}\right)^{-1} \mathbf{R} \right)}_{\mathbf{S}_0} \mathbf{a} 
\end{equation}
 is obtained.  Since, by definition, $\mathbf{\Gamma}_\mathrm{L,0}$ is the termination to obtain the scattering object defined by $\mathbf{S}_0$, it is assumed that the terminated antenna should behave identically to the scattering object from the characteristic mode analysis. Therefore, the relationship between $\mathbf{S}_0$ and $\mathbf{S}$ is defined by the following equation:
\begin{equation}
    \label{eq:gsm_from_sm_without_losslessness_relations_applied}
    \mathbf{S} = \mathbf{S}_0 - \mathbf{T}  \left(\mathbf{\Gamma}_\mathrm{L,0}-\mathbf{\Gamma}\right)^{-1} \mathbf{R}.
\end{equation}

%Since the antenna is defined to have uncoupled and matched ports, each port is independent. This also means that the ports can be terminated by individual reactive loads $|\Gamma_{\mathrm{L},0,p}| = 1$, making $\mathbf{\Gamma}_\mathrm{L,0}$ a diagonal unitary matrix.

%\subsection{Special Case for Lossless Reciprocal Matched Antennas }

Now, it is shown for a specific type of antennas, that if the eigenvalues $\lambda_n$ of the terminated antenna and the antenna transmit matrix $\mathbf{T}$ are known, the scattering matrix $\mathbf{S}$ of the antenna is also defined implicitly. This type of antennas are antennas which are reciprocal, lossless, matched and have decoupled ports:
\begin{equation}
\label{eq:gsm_def_scat_synth}
    \mathbf{\Psi} = \begin{bmatrix}
  {\mathbf{S}}&{\mathbf{T}} \\ 
  {\mathbf{T}^\mathrm{T}}&\mathbf{0} 
\end{bmatrix} \hspace{3mm}\text{where}\hspace{3mm} {\mathbf{S}} = \mathbf{S}^\mathrm{T}. %\hspace{2mm}\text{and}\hspace{2mm}\mathbf{T}^\mathrm{T} \mathbf{T}^* = \mathbf{I}.
\end{equation}

As this type of antennas is lossless by definition, the following equations apply:
\begin{equation}
    \label{eq:losslessness_unitary_of_t}
    \mathbf{T}^\mathrm{T} \mathbf{T}^* = \mathbf{I},
\end{equation}
\begin{equation}
    \label{eq:losslessness_orthogonality_of_S_and_conj_T}
    \mathbf{S} \mathbf{T}^* = \mathbf{0}.
\end{equation}

Inserting \eqref{eq:gsm_from_sm_without_losslessness_relations_applied} and \eqref{eq:losslessness_unitary_of_t} into \eqref{eq:losslessness_orthogonality_of_S_and_conj_T} yields
\begin{equation}
    \label{eq:condition_for_T}
    \mathbf{S}_0 \mathbf{T}^* = \mathbf{T} \mathbf{\Gamma}_\mathrm{L,0}^{-1}.
\end{equation}

%This equation can be interpreted in the following way: For $\mathbf{T}$ to be a valid antenna transmit matrix of a lossless and reciprocal antenna with uncoupled ports that originates from a scattering object with scattering matrix $\mathbf{S}_0$, there must exist a diagonal unitary matrix $\mathbf{\Gamma}_\mathrm{L,0}$ such that \eqref{eq:condition_for_T} is fulfilled.

Combining this equation together with \eqref{eq:gsm_from_sm_without_losslessness_relations_applied} leads to
\begin{equation}
\label{eq:gsm_from_sm}
    \mathbf{S} = \mathbf{S}_0\left(\mathbf{I}- \mathbf{T}^*\mathbf{T}^\mathrm{T}\right),
\end{equation}
which is independent of $\mathbf{\Gamma}_\mathrm{L,0}$.  This equation describes the desired relation  by which the scattering matrix of the antenna $\mathbf{S}$ can be described based on the radiation behavior of the antenna (represented by $\mathbf{T}$) and the eigenvalues of the element geometry (represented by $\mathbf{S}_0$). 

%\annotateChanges{It represents a description of the scattering matrix $\mathbf{S}$ of an antenna in terms of the degrees of freedom of the antenna, whereby $\mathbf{T}$ describes the radiation behavior of the antenna and $\mathbf{S}_0$ defines the residual scattering behavior of the antenna that is 'not connected` to a port.
 
It is noted that a special case of the proposed formula \eqref{eq:gsm_from_sm} is obtained when the reference scattering object is assumed to be free-space ($\mathbf{S}_0 = \mathbf{I}$):
\begin{equation}
\label{eq:ms_antenna_definition}
    \mathbf{S} = \mathbf{I} - \mathbf{T}^* \mathbf{T}^\mathrm{T}.
\end{equation}
This formula is known from \cite{Kahn1965} to construct the scattering matrix of a minimum scattering antenna\footnote{ Note \cite[eq. (12)]{Kahn1965} is a variant of \eqref{eq:ms_antenna_definition} for the more specific case of a (reciprocal) canonical minimum scattering antenna where $\mathbf{T} = \mathbf{T}^*$. Here, the more general case of a (reciprocal) minimum scattering antenna is assumed.{}}. Formally, a minimum scattering antenna describes an antenna for which a set of lossless port terminations exists such that the antenna becomes electromagnetically invisible. Now, with \eqref{eq:gsm_from_sm} on the other hand, the general case that includes both minimum scattering antennas and non-minimum scattering antennas is covered.

A proof that \eqref{eq:gsm_from_sm} describes a lossless antenna is given in Appendix~\ref{appendix:scatterer_to_antenna}.

\subsection{Synthesis of Antennas in Terms of Generalized Scattering Matrices}
\label{sec:synthesis_approach}

Now, to enable the mathematical synthesis of antenna elements, a model of the antenna elements with degrees of freedom is developed, as depicted in Fig.~\ref{fig:overview_synthesis}.

Usually, the degrees of freedom of antenna elements are geometrical parameters like e.g. widths, lengths, excitation positions and others. However, the main idea of the proposed synthesis approach is that the variation of these geometrical parameters can also be represented by a variation of abstract parameters that describe the radiation and scattering behavior of the elements.

For the description of these parameters, a generalized scattering matrix in terms of the characteristic modes of a structure that is close to the final antenna element is used.

%Therefore, the (later justified) assumption is that the element geometry and its geometric parameters remain fixed for the entire mathematical synthesis and variation is introduced by changing only abstract modal parameters.

Therefore, in a first step, an initial element geometry is specified without defining ports.
Then, the characteristic modes are calculated for this element geometry.
The resulting eigenfields $\mathbf{E}_n$ become the basis functions of the generalized scattering matrix description of the synthesized element. The resulting eigenvalues $\lambda_n$ however, are not used for the synthesis.
Instead, to describe all antennas with an element geometry close to the initial element geometry, modal degrees of freedom are introduced.
%Now, instead of placing actual ports into the element geometry, abstract synthetic ports are introduced based on a composition of the eigencurrents.

These degrees of freedom are the synthetic antenna transmit matrix $\mathbf{T}^\prime$ and the synthetic eigenvalues $\lambda_n^\prime$, whereby the prime~$^\prime$ is introduced to indicate that these parameters are synthetic.
Using \eqref{eq:gsm_from_sm}, the synthetic scattering matrix of the antenna $\mathbf{S}^\prime$ is then defined based on $\mathbf{S}_0^\prime$ and $\mathbf{T}^\prime$, which completes the generalized scattering matrix of the antenna element.

%This provides the mathematical model of the generalized scattering matrix of the antenna element in terms of the abstract modal degrees of freedom $\lambda_n^\prime$ and $\mathbf{T}^\prime$.

This parameterized generalized scattering matrix can now be used as a basis for optimization to find the modal degrees of freedom that yield e.g. optimal mutual coupling, optimal patterns or other optimization goals.
%To find an implementation of this synthetic antenna element, the synthesized eigenvalues $\lambda_n^\prime$ and the synthesized antenna transmit matrix $\mathbf{T}^\prime$ have to be realized.

Once the optimization goals are met, an actual antenna element can be designed that has the particular modal configuration $\lambda_n^\prime$ and $\mathbf{T}^\prime$ that was found during the optimization.
Here, it is noted that especially the eigenvalues $\lambda_n$ depend on the investigated element geometry.
So, in order to realize this antenna element, the initial element geometry has to be modified.

Strictly, this new element geometry has different eigenfields than the initial element geometry from the synthesis.
\annotateChangesStart However, if the element geometry is changed in a controlled manner such that the eigenfields $\mathbf{E}_n$ only change marginally at a certain distance from the structure, this effect can be neglected.\annotateChangesEnd{}
%, so that the effect of the change of the eigencurrents $\mathbf{J}_n$ can be considered negligible.
Depending on the element geometry, this distance can be already reached within the near-field of the element such that the proposed approach can be used to synthesize elements within an array as done later in this manuscript.

%as far as the application within antenna arrays later in this paper is considered and

% There are, in fact, several ways to control of the eigenvalues $\lambda_n$ of the structure during the design of the structure \cite{Adams2022}.
% Because of this, eigenvalues serve as excellent degrees of freedom for the mathematical synthesis of an antenna.
% Therefore, the eigenvalues from the calculation of the characteristic modes $\lambda_n$ are replaced by synthetic degrees of freedom $\lambda_n^\prime$.

% Furthermore, the entries of the antenna transmit matrix $\mathbf{T}$ of the antenna are used as degrees of freedom.

%so that the change in current densities can be neglected in order to describe the scattering behaviour of the structure.
%This means that the 

%In the design of antennas using characteristic modes, the 

\subsection{Reduction of the Degrees of Freedom}

In terms of reciprocity, it is noted, that according to \eqref{eq:gsm_from_sm}, ${\mathbf{S}}$ is not always guaranteed to be symmetric for any $\mathbf{T}$ that fulfills \eqref{eq:losslessness_unitary_of_t}. Indeed, ${\mathbf{S}}$ is symmetric if and only if
\begin{equation}
    \label{eq:condition_formulation2}
    \mathbf{S}_0\,\mathbf{T}^{*}\mathbf{T}^{{\text{T}}} - {\mathbf{T}}\mathbf{T}^{{\text{H}}}\mathbf{S}_0\, = \mathbf{0}.
\end{equation}
%which is kind of similar to a zero commutator\footnote{When $\mathbf{A}=\mathbf{A}^{\mathrm{T}}$ and  $\mathbf{B}=\mathbf{B}^{\mathrm{H}}$, then $\mathbf{A}\mathbf{B}$ is symmetric if and only if $\mathbf{A}\mathbf{B}-\mathbf{B}^*\mathbf{A} = \mathbf{0}$. Here, $\mathbf{A}\mathbf{B}-\mathbf{B}^*\mathbf{A} = \mathbf{0}$ applies, whereby a zero commutator would mean: $\mathbf{A}\mathbf{B}-\mathbf{B}\mathbf{A} = \mathbf{0}$.}.
This implies that the conversion from a scattering object $\mathbf{S}_0$ to a reciprocal antenna is only valid for certain $\mathbf{T}$. Consequently, the degrees of freedom for the synthesis are governed by $\mathbf{S}_0$ and $\mathbf{T}$, yet not all entries are individual degrees of freedom since \eqref{eq:condition_formulation2} establishes a relation between $\mathbf{S}_0$ and $\mathbf{T}$. This has to be accounted for during the formulation of the degrees of freedom of the antenna. How this can be done, is shown for an example in section~\ref{sec:synth_model}.

\section{Considerations on the Array Level}
\label{sec:theory_array_level}

\begin{figure}
\centering
\begin{tikzpicture}
    \pgfmathsetmacro{\boxWidth}{2}
    \pgfmathsetmacro{\arrowDist}{0.3}
    \pgfmathsetmacro{\arrowLength}{1.2}
    \pgfmathsetmacro{\smallBoxDist}{0.4}
    \pgfmathsetmacro{\smallBoxWidth}{1.1cm}
    
    \node [anchor=north east,draw,minimum height=\smallBoxWidth,minimum width=\smallBoxWidth] (box0) at (-\arrowLength, 0) {$\mathbf{\Psi}^{(1)}$};
    \node [below=\smallBoxDist of box0.south,draw,minimum height=\smallBoxWidth,minimum width=\smallBoxWidth] (box1) {$\mathbf{\Psi}^{(2)}$};
    \node [below=\smallBoxDist of box1.south] (box2) {$\cdots$};
    \node [below=\smallBoxDist of box2.south,draw,minimum height=\smallBoxWidth,minimum width=\smallBoxWidth] (box3) {$\mathbf{\Psi}^{(K)}$};
    
    \def\arrowDir{->}
    \newcommand{\makeArrows}[1]{
        \foreach \x in {0,1,3} {
            \draw[\arrowDir] ($ (box\x.#1) + (\arrowOffset, 0.5*\arrowDist) $) -- +(\arrowLength, 0);
            \draw[\arrowDir] ($ (box\x.#1) + (\arrowLength+\arrowOffset, -0.5*\arrowDist) $) -- +(-\arrowLength, 0);
        }
    }

    \pgfmathsetmacro{\arrowOffset}{0}
    \makeArrows{east}
    
    \coordinate[left=0.5*\arrowLength of box0.west] (a);
    \node[above=10mm of a, text width=1.8cm,align=center] (portif) {Port Interface ($\mathbf{v}^{(k)}$, $\mathbf{w}^{(k)}$)};
    \draw[dashed] (portif.south) -- (portif.south |- box3.south east);
    
    \coordinate[right=0.5*\arrowLength of box0.east] (a);
    \node[above=10mm of a, text width=2cm,align=center] (portif) {Radiation Interface ($\mathbf{a}^{(k)}$, $\mathbf{f}^{(k)}$)};
    
    \draw[dashed] (portif.south) -- (portif.south |- box3.south east);
    
    \coordinate (start) at (0, 0);
    
    \pgfmathsetmacro{\arrowOffset}{-\arrowLength}
    \makeArrows{west}
    
    \coordinate (bottomLeft) at (start |- box3.south east);
    \coordinate (bottomCenter) at ($(bottomLeft) + (0.5*\boxWidth, 0)$);
    
    \draw[] (0,0) -- +(\boxWidth,0) |- (bottomLeft) -- cycle;
    \node at ($(start |- box3.south east)!0.5!(\boxWidth,0)$) {$\mathbf{G}$};
    
    \pgfmathsetmacro{\labelOffset}{\arrowLength+0.3}
    
    \node[left=\labelOffset of box0.west,anchor=east] {$1^{\text{st}}$ Element:};
    \node[left=\labelOffset of box1.west,anchor=east] (lastLabel) {$2^{\text{nd}}$ Element:};
    \node at (lastLabel.south |- box2.west)  {$\cdots$};
    \node[left=\labelOffset of box3.west,anchor=east] {$K^{\text{th}}$ Element:};
    
    \node[below=5mm of box3.south,anchor=north,text width=3cm,align=center] (lastLabel) {Generalized Scattering Matrices of the Array Elements};
    \node[text width=2cm,align=center] at (bottomCenter |- lastLabel.east) {Modal Coupling Matrix};
    
\end{tikzpicture}
\caption{Visualization of the interaction between $K$ antenna elements in an array, represented by the modal coupling matrix $\mathbf{G}$ and the generalized scattering matrices $\mathbf{\Psi}^{(k)}$ of the individual array elements.}
\label{fig:gsm_coupling}
\end{figure}
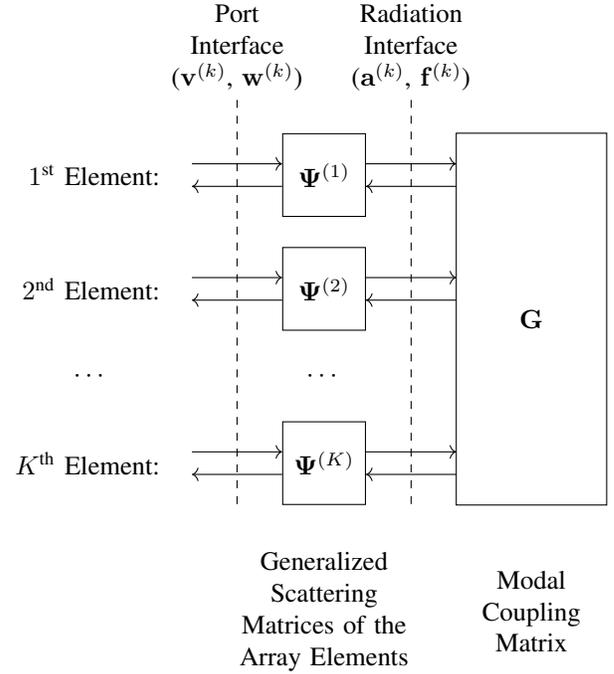
%\subsection{Array Calculations Using the Generalized Scattering Matrices of the Array Elements}

Now, after an approach for the synthesis of isolated elements has been introduced, the attention is shifted towards the array level. To pick up on the mathematical description from the previous section, the array is thereby composed of elements represented by their generalized scattering matrices and a modal coupling matrix is introduced to connect them. This is illustrated in Fig.~\ref{fig:gsm_coupling}.

%In the following, a method that connects the description of the isolated elements from the last section into an array is discussed. The 

%The goal of the proposed model is to allow the decomposition of the mutual coupling in antenna arrays into modes to enhance the understanding of coupling phenomena. Thereby, it should be noted that even in the transmitting case, the antenna elements in an array act simultaneously as radiators and as scatterers since each element is also illuminated by adjacent antenna elements.

In the following, relationships to describe the mutual coupling between the array elements in this formalism are derived. The derivation is inspired by \cite{Rubio2003,Rubio2005}, where Rubio et al. discussed similar relationships in the basis of spherical wave functions.

First, at every antenna element $k$, the generalized scattering matrix $\mathbf{\Psi}^{(k)}$ is described by:
\begin{equation}
\label{eq:gsm_generic_element_k}
    \begin{bmatrix}
   {\mathbf{S}^{(k)}}-\mathbf{I}&{\mathbf{T}^{(k)}} \\ 
  {\mathbf{R}^{(k)}}&{\mathbf{\Gamma }^{(k)}} 
\end{bmatrix}\begin{bmatrix}
  {\mathbf{a}^{(k)}} \\ 
  {\mathbf{v}^{(k)}} 
\end{bmatrix} = \begin{bmatrix}
  {\mathbf{f}^{(k)}} \\ 
  {\mathbf{w}^{(k)}} 
\end{bmatrix},
\end{equation}
where $\mathbf{f}^{(k)} = \mathbf{b}^{(k)} - \mathbf{a}^{(k)}$ is used on the right-hand side in contrast to \eqref{eq:gsm_generic}. Since ${\mathbf{S}^{(k)}}-\mathbf{I}$ is used in the generalized scattering matrix on the left-hand side, the description remains identical.

These generalized scattering matrices $\mathbf{\Psi}^{(k)}$ are coupled to each other because the incident field at each antenna $k$ includes a component resulting from the current distribution on adjacent antenna elements and a component due to the field externally incident onto the array. In terms of the modal weighting coefficients, this is written as:
\begin{equation}
    {\mathbf{a}^{(k)}} = {\mathbf{a}_{\mathrm{ext}}^{(k)}} + \sum_{\substack{l=1 \\ l\neq k}}^{K} {\mathbf{a}^{(k\leftarrow l)}},
\end{equation}
whereby $ {\mathbf{a}_{\mathrm{ext}}^{(k)}}$ contains the modal weighting coefficients of the external incident field and ${\mathbf{a}^{(k\leftarrow l)}}$ contains the modal weighting coefficients of the incident field due to coupling from the $l$\nobreakdash-th to the $k$\nobreakdash-th element.

To establish the connection to the weighting coefficients of the outgoing modes on the adjacent antenna elements $\mathbf{f}^{(l)}$, the weighting coefficients due to coupling ${\mathbf{a}^{(k\leftarrow l)}}$ can be described using:
\begin{equation}
     {\mathbf{a}^{(k\leftarrow l)}} = \mathbf{G}^{(k,l)} \mathbf{f}^{(l)}.
\end{equation}
Here, $\mathbf{G}^{(k,l)}$ is the modal coupling matrix between the modes of the $k$\nobreakdash-th element and the $l$\nobreakdash-th element. As derived in Appedix~\ref{sec:G_calculation}, if an electric field integral equation scheme of the method of moments is considered, the modal coupling matrix can be obtained in the following way:
\begin{equation}
\label{eq:G_coupling_CM}
    \mathbf{G}^{(k,l)} = \frac{1}{2} {\mathbf{I}^{(k)\mathrm{T}}_{\mathrm{CM}}} \mathbf{Z}^{(k,l)} {\mathbf{I}^{(l)}_{\mathrm{CM}}}.
\end{equation}
Thereby, $\mathbf{Z}^{(k,l)}$ is the submatrix block of the impedance matrix $\mathbf{Z}$ that describes coupling from the basis functions of the $l$\nobreakdash-th array element to the testing functions on the $k$\nobreakdash-th array element. ${\mathbf{I}^{(k)}_{\mathrm{CM}}}$ and ${\mathbf{I}^{(l)}_{\mathrm{CM}}}$ are matrices whose columns contain the characteristic modes on the $k$\nobreakdash-th element and the $l$\nobreakdash-th element, each normalized such that it radiates $\SI{0.5}{\watt}$ as it is usual for scattering matrix analyses \cite{Hansen1988}.

In total, this leads to a system of $K$ coupled generalized scattering matrices:
\begin{align}
\label{eq:system_of_K_gsms}
\begin{split}
    \mathbf{\Gamma}^{(k)} \mathbf{v}^{(k)} + \mathbf{R}^{(k)} {\mathbf{a}_{\mathrm{ext}}^{(k)}} + \mathbf{R}^{(k)} \sum_{\substack{l=1 \\ l\neq k}}^{K} \mathbf{G}^{(k,l)} \mathbf{f}^{(l)} &= \mathbf{w}^{(k)}, \\
    \mathbf{T}^{(k)} \mathbf{v}^{(k)} + \left(\mathbf{S}^{(k)} - \mathbf{I}\right) \left( {\mathbf{a}_{\mathrm{ext}}^{(k)}} + \sum_{\substack{l=1 \\ l\neq k}}^{K} \mathbf{G}^{(k,l)} \mathbf{f}^{(l)} \right) &= \mathbf{f}^{(k)}.
\end{split}
\end{align}

%While Rubio et al. use spherical wave functions as modes at the radiation interface, generalized scattering matrices can also be defined with respect to other modal decompositions at the radiation interface. In this work, the generalized scattering matrices are formulated in terms of characteristic modes. How these generalized scattering matrices can be calculated, is explained in subsection~\ref{sec:gsm_cm_directly_from_mom}. For now, it is assumed that they are already given and consider how to formulate the coupling between them.

Similar to Rubio et al., block diagonal matrices are introduced that contain the parts of the  generalized scattering matrices of the isolated elements:
\begin{equation}
\begin{aligned}
    {{\mathbf{S}}^{{\text{(iso)}}}} = {\text{diag}}\left\{ {{{\mathbf{S}}^{(k)}}} \right\};& \hspace{3mm} {{\mathbf{T}}^{{\text{(iso)}}}} = {\text{diag}}\left\{ {{{\mathbf{T}}^{(k)}}} \right\}; \\
    {{\mathbf{R}}^{{\text{(iso)}}}} = {\text{diag}}\left\{ {{{\mathbf{R}}^{(k)}}} \right\};& \hspace{3mm} {{\mathbf{\Gamma }}^{{\text{(iso)}}}} = {\text{diag}}\left\{ {{{\mathbf{\Gamma }}^{(k)}}} \right\}.
\end{aligned}
\end{equation}

 Also, all coupling matrices $\mathbf{G}^{(k,l)} $ are collected in a modal coupling matrix ${\mathbf{G}}$:
\begin{equation}
{\mathbf{G}} = \begin{bmatrix}
  {\mathbf{0}}&{{{\mathbf{G}}^{(1,2)}}}&{...}&{{{\mathbf{G}}^{(1,K)}}} \\ 
  {{{\mathbf{G}}^{(2,1)}}}&{\mathbf{0}}& \ddots & \vdots  \\ 
   \vdots & \ddots & \ddots &{{{\mathbf{G}}^{(K - 1,K)}}} \\ 
  {{{\mathbf{G}}^{(K,1)}}}&{...}&{{{\mathbf{G}}^{(K,K - 1)}}}&{\mathbf{0}} 
\end{bmatrix},
\end{equation}
 and all incident and scattered wave coefficients are also written in a compact form:
\begin{equation}
\label{eq:block_diagonal_matrices_definition}
\begin{aligned}
    \mathbf{v} = \begin{bmatrix}\mathbf{v}^{(1)}\\\vdots\\\mathbf{v}^{(K)}\end{bmatrix} &;\hspace{3mm}
    \mathbf{w} = \begin{bmatrix}\mathbf{w}^{(1)}\\\vdots\\\mathbf{w}^{(K)}\end{bmatrix}; \\
    \mathbf{a}_{\mathrm{ext}} = \begin{bmatrix}\mathbf{a}^{(1)}_{\mathrm{ext}}\\\vdots\\\mathbf{a}^{(K)}_{\mathrm{ext}}\end{bmatrix} &;\hspace{3mm}
    \mathbf{f} = \begin{bmatrix}\mathbf{f}^{(1)}\\\vdots\\\mathbf{f}^{(K)}\end{bmatrix}.
\end{aligned}
\end{equation}

Now, using this notation, the system of equations from \eqref{eq:system_of_K_gsms} can be rewritten as a new generalized scattering matrix that contains coupling between the elements:
\begin{equation}
    \begin{bmatrix}
   {\mathbf{S}^{\text{(coupled)}}}-\mathbf{I}&{\mathbf{T}^{\text{(coupled)}}} \\ 
  {\mathbf{R}^{\text{(coupled)}}}&{\mathbf{\Gamma }^{\text{(coupled)}} }
\end{bmatrix}\begin{bmatrix}
  \mathbf{a}_{\text{ext}} \\ 
  \mathbf{v} 
\end{bmatrix} = \begin{bmatrix}
  \mathbf{f} \\ 
  \mathbf{w} 
\end{bmatrix}.
\end{equation}
The coupled matrices are given by:
\begin{equation}
\label{eq:arr_gsm}
\begin{aligned}
    {{\mathbf{\Gamma }}^{{\text{(coupled)}}}} &= {{\mathbf{\Gamma }}^{{\text{(iso)}}}} + {{\mathbf{R}}^{{\text{(iso)}}}}{\mathbf{G}}\,{\mathbf{M}}\,{{\mathbf{T}}^{{\text{(iso)}}}} \\
    {{\mathbf{R}}^{{\text{(coupled)}}}} &= {{\mathbf{R}}^{{\text{(iso)}}}} + {{\mathbf{R}}^{{\text{(iso)}}}}{\mathbf{GM}}\left( {{{\mathbf{S}}^{{\text{(iso)}}}} - {\mathbf{I}}} \right) \\
    {{\mathbf{T}}^{{\text{(coupled)}}}} &= {\mathbf{M}}\,{{\mathbf{T}}^{{\text{(iso)}}}} \\
    \left( {{{\mathbf{S}}^{{\text{(coupled)}}}} - {\mathbf{I}}} \right) &= {\mathbf{M}}\left( {{{\mathbf{S}}^{{\text{(iso)}}}} - {\mathbf{I}}} \right)
\end{aligned}
\end{equation}
whereby
\begin{equation}
    {\mathbf{M}} = {\left( {{\mathbf{I}} - \left( {{{\mathbf{S}}^{{\text{(iso)}}}} - {\mathbf{I}}} \right){\mathbf{G}}\,} \right)^{ - 1}}.
\end{equation}
The matrix ${{\mathbf{\Gamma }}^{{\text{(coupled)}}}}$ contains the port scattering parameters of all elements in the antenna array. The matrix ${{\mathbf{T}}^{{\text{(coupled)}}}}$ is called the embedded antenna transmit matrix following the term `embedded element pattern'. The terms ${{\mathbf{R}}^{{\text{(coupled)}}}}$ and ${{\mathbf{S}}^{{\text{(coupled)}}}}$ describe the receive and scattering behavior of the array when external incident waves exist.

When the current distribution on the array is of interest, it can be calculated on the $k$-th element using:
\begin{equation}
\label{eq:Ip_excited}
    \mathbf{I}^{(k)} = \mathbf{I}^{(k)}_{\mathrm{CM}} \,\mathbf{f}^{(k)},
\end{equation}
and the total array far-field pattern $\mathbf{F}(\theta, \phi)$ can be either calculated from these currents or analogously based on a superposition of the far-fields of the characteristic modes:
\begin{equation}
\label{eq:Fp_excited}
    \mathbf{F}(\theta, \phi) = \sum_{k = 1}^{K}  \sum_{n = 1}^{N} \mathbf{F}_{\mathrm{CM},n}^{(k)}(\theta, \phi) f^{(k)}_{n},
\end{equation}
whereby $\mathbf{F}_{\mathrm{CM},n}^{(k)}(\theta, \phi)$ is the far-field of the $n$\nobreakdash-th characteristic mode on the $k$\nobreakdash-th element $\mathbf{I}^{(k)}_{\mathrm{CM},n}$. The index $n$ thereby denotes the $n$\nobreakdash-th column of $\mathbf{I}^{(k)}_{\mathrm{CM}}$.

\section{Example: Synthesis of a Circularly Polarized Patch Antenna Array}
\label{sec:example}

\begin{figure}
    \centering
    \includegraphics{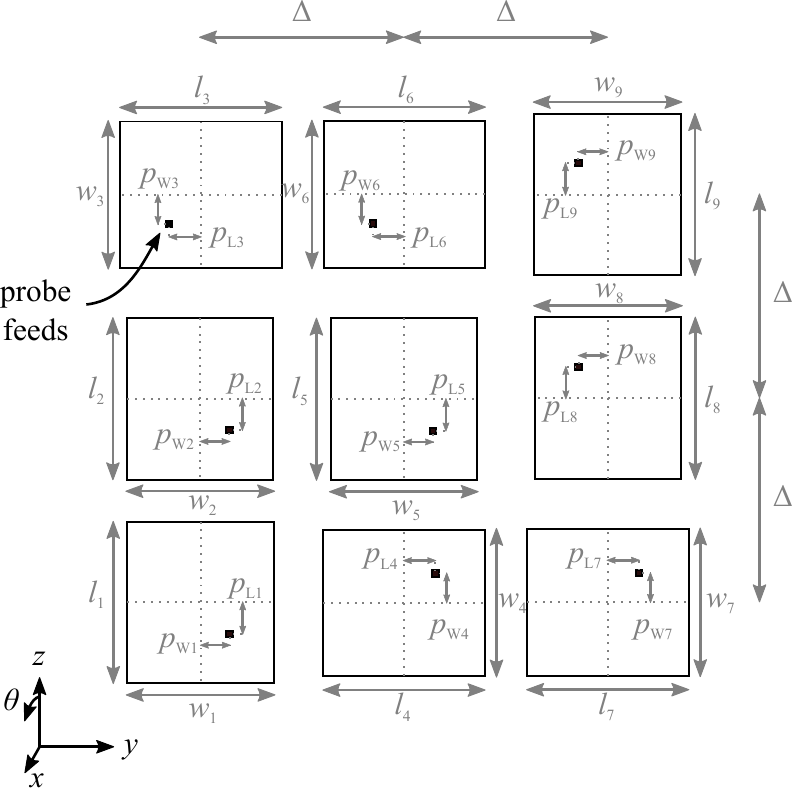}
    \caption{Geometry of the $3 \times 3$ LHCP patch antenna array.}
    \label{fig:circular_patch_array_sketch}
\end{figure}

%\subsection{Degrees of Freedom in the Synthesis of Reciprocal Antennas}
%\label{subsection:dof_of_synthesis}
Now that a formulation of synthetic array elements and their coupling has been given, an example is used to show how a design procedure for an array can be derived on this basis.
Therefore, exemplary, a left-handed circularly polarized (LHCP) patch antenna array with high cross-polarization rejection and a fixed beam is synthesized in this section.

The investigation focuses on a $3 \times 3$ LHCP patch antenna array, as depicted in Fig.~\ref{fig:circular_patch_array_sketch}.
The array is supposed to radiate LHCP at $f_0 = \SI{28}{\giga\hertz}$ in the $\theta$-plane and is built $\lambda_0/20$ in front of an infinite ground plane that is parallel to the $yz$-plane. The inter-element distance is $\Delta = 0.56\lambda_0$, the patches are modeled as perfectly conducting, infinitely thin sheets and the probe feeds have a width of $\SI{200}{\micro\meter}$. The edge-lengths of the patches $w_k$ and $l_k$ and the positions of the probe feeds $p_{\mathrm{W},k}$ and $p_{\mathrm{L},k}$ are  optimized in the following to achieve a high cross-polarization rejection.

\begin{figure*}
    \centering
    \includegraphics{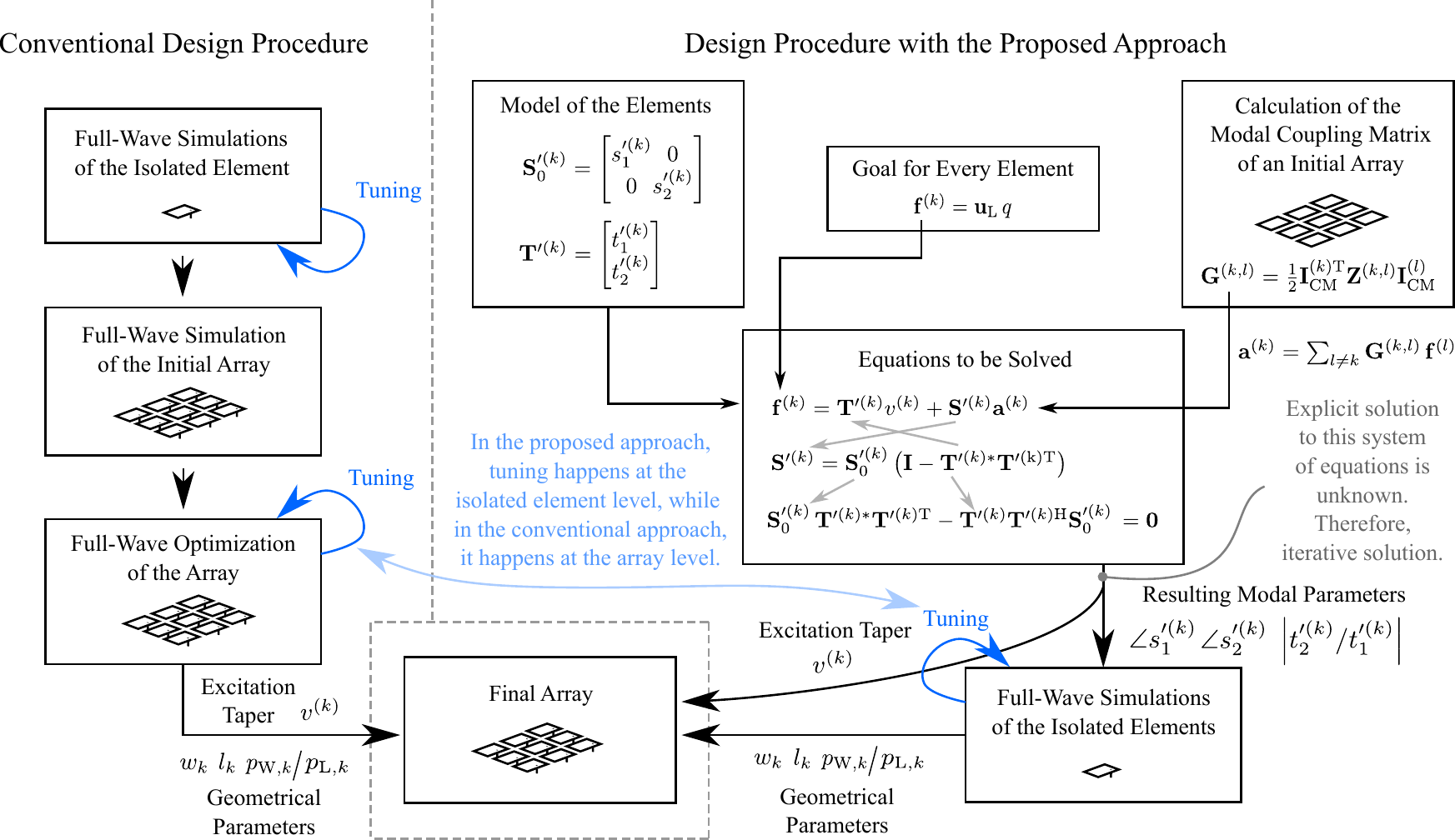}
    \caption{ Comparison of the design procedures for the design of a $3 \times 3$ circularly polarized patch array using the conventional approach and using the proposed approach.}
    \label{fig:overview_example}
\end{figure*}

On the one hand, such a design task can be solved using a conventional design procedure, where the performance of the array is tuned using repeated full-wave simulations of the entire array. In this approach, conclusions about the design parameters are achieved indirectly through experience or through random variation. On the other hand, the proposed approach can be used, where a mathematical model based on physical relationships is established that allows direct conclusions about the relationships between the design parameters and the desired results. In Fig.~\ref{fig:overview_example}, the two approaches are compared to each other.

\subsection{First Steps of the Conventional Design Procedure}

For reference, the first steps of the conventional design procedure (as shown in Fig.~\ref{fig:overview_example}) are carried out. The procedure starts with the tuning of a single isolated element, such that only LHCP is radiated broadside. This is achieved using full-wave simulations of the isolated element. The resulting geometric parameters of the tuned element are: $w_k = \SI{4.3}{\milli\meter}$, $l_k = \SI{4.75}{\milli\meter}$, $p_{\mathrm{W},k} = \SI{850}{\micro\meter}$, $p_{\mathrm{L},k} = \SI{900}{\micro\meter}$.

Next, an initial array is set up, where these geometric parameters are used for all elements in the array configuration. 
 As an additional measure to achieve a high cross-polarization rejection, the outer elements of the array are arranged according to the sequential rotation principle \cite{Teshirogi1985}. According to this principle, the same element is placed in different orientations in the array so that the undesired circular cross-polarization radiation of the elements cancels each other (theoretically) independently of the polarizations of the element radiation patterns. This fact makes these arrays typically resistant to mutual coupling effects. However, the sequential rotation principle cannot be applied to the central element of this array since it has no `rotation partner' that cancels the cross-polarization radiation components.  Therefore, it will be seen in the following, that the undesired cross-polarization radiation due to mutual coupling has to be suppressed with additional measures in this case.

% \setlength{\tabcolsep}{0.4em}
% \renewcommand{\arraystretch}{1.2}
% \begin{table}
%  \caption{Geometric Parameters of the Initial circularly Polarized patch Array}
% \label{tab:circ_patch_geometry_init}
% \begin{tabular}{|c*{9}{|>{\centering\arraybackslash}p{0.55cm}}|}
% \hline
% \rule{0pt}{15pt}
% $k$
% \rule{0pt}{15pt}& 1    & 2    & 3    & 4    & 5    & 6    & 7    & 8    & 9    \\[5pt] \hline \hline
% \begin{tabular}[c]{@{}c@{}}$w_k$\\ in mm\end{tabular}           & 4.3  & 4.3  & 4.3  & 4.3  & 4.3  & 4.3  & 4.3  & 4.3  & 4.3  \\ \hline
% \begin{tabular}[c]{@{}c@{}}$l_k$\\ in mm\end{tabular}           & 4.75 & 4.75 & 4.75 & 4.75 & 4.75 & 4.75 & 4.75 & 4.75 & 4.75 \\ \hline
% \begin{tabular}[c]{@{}c@{}}$p_{\text{W}k}$\\ in \si{\micro\meter}\end{tabular} & 850  & 850  & 850  & 850  & 850  & 850  & 850  & 850  & 850  \\ \hline
% \begin{tabular}[c]{@{}c@{}}$p_{\text{L}k}$\\ in \si{\micro\meter}\end{tabular} & 900  & 900  & 900  & 900  & 900  & 900  & 900  & 900  & 900  \\ \hline
% \end{tabular}
% \end{table}

\begin{figure}
    \centering
\definecolor{mycolor1}{rgb}{0.65098,0.80784,0.89020}%
\definecolor{mycolor2}{rgb}{0.12157,0.47059,0.70588}%
\definecolor{mycolor4}{rgb}{1.00000,0.49804,0.00000}%
\definecolor{mycolor3}{rgb}{0.41569,0.23922,0.60392}%
\begin{tikzpicture}

\begin{axis}[%
width=2.441in,
height=1.743in,
at={(0.409in,0.442in)},
scale only axis,
xmin=0,
xmax=180,
xtick={  0,  30,  60,  90, 120, 150, 180},
xlabel style={font=\color{white!15!black}},
xlabel={$\theta\text{ in }{^\circ}$},
ylabel={Realized gain in dBi},
ymin=-20,
ymax=20,
ytick={-25, -20, -15, -10,  -5,   0,   5,  10,  15,  20,  25},
axis background/.style={fill=white},
xmajorgrids,
ymajorgrids
]
\addplot [color=mycolor2, line width=1.0pt, forget plot]
  table[row sep=crcr]{%
0	-18.2958861993841\\
1	-17.3207420401622\\
2	-16.3985452161468\\
3	-15.5202086004177\\
4	-14.6787189149409\\
5	-13.8686359316538\\
6	-13.0857289513242\\
7	-12.3267086352607\\
8	-11.5890259656164\\
9	-10.8707192340239\\
10	-10.1702959930004\\
11	-9.48664091059685\\
12	-8.81894315071532\\
13	-8.16663871662956\\
14	-7.52936443890863\\
15	-6.90692115268624\\
16	-6.29924421807786\\
17	-5.70637997382766\\
18	-5.12846703257761\\
19	-4.56572156301237\\
20	-4.01842588416379\\
21	-3.48691983710001\\
22	-2.97159451059578\\
23	-2.47288798822762\\
24	-1.99128286041973\\
25	-1.52730531054584\\
26	-1.08152564258492\\
27	-0.654560171795611\\
28	-0.247074451908263\\
29	0.140212135094961\\
30	0.506520344765907\\
31	0.851003447601498\\
32	1.17273780187433\\
33	1.47071131824\\
34	1.74380940316308\\
35	1.99079782775927\\
36	2.21030178404089\\
37	2.40078014303741\\
38	2.56049359141447\\
39	2.68746485416426\\
40	2.7794285486568\\
41	2.83376726384247\\
42	2.8474290660535\\
43	2.81681955467706\\
44	2.73765842124431\\
45	2.60478551368613\\
46	2.41189346216815\\
47	2.15115078475156\\
48	1.81265691059978\\
49	1.38363055422339\\
50	0.847158377761989\\
51	0.180184559329564\\
52	-0.649884282806677\\
53	-1.69127586337123\\
54	-3.01968286464873\\
55	-4.76363713322842\\
56	-7.16656049337604\\
57	-10.7643058987059\\
58	-16.7820774329951\\
59	-18.0631039557006\\
60	-10.9404400299518\\
61	-6.40785237777957\\
62	-3.2520692896719\\
63	-0.831625603928999\\
64	1.13513237414603\\
65	2.79281145747405\\
66	4.22482700773685\\
67	5.48347749579391\\
68	6.60363560025189\\
69	7.60967585571657\\
70	8.51927747998666\\
71	9.34565040240822\\
72	10.0989076297478\\
73	10.7869481878042\\
74	11.4160458012761\\
75	11.9912534154376\\
76	12.5166884469458\\
77	12.9957384609686\\
78	13.4312123600603\\
79	13.8254533912939\\
80	14.1804248380909\\
81	14.4977757983463\\
82	14.7788921886958\\
83	15.0249366050712\\
84	15.2368796411095\\
85	15.4155245515746\\
86	15.5615266424619\\
87	15.6754084050249\\
88	15.7575711429543\\
89	15.808303640325\\
90	15.8277882620574\\
91	15.8161047538894\\
92	15.7732319044842\\
93	15.6990471399784\\
94	15.5933240340719\\
95	15.4557276282972\\
96	15.2858073608819\\
97	15.082987291226\\
98	14.846553171294\\
99	14.575635743036\\
100	14.2691894155118\\
101	13.9259651726484\\
102	13.5444761471814\\
103	13.1229537152167\\
104	12.6592911374891\\
105	12.1509705699544\\
106	11.5949674827341\\
107	10.9876238254351\\
108	10.3244770889306\\
109	9.60002574753005\\
110	8.80740063685732\\
111	7.93789329551186\\
112	6.98025965709257\\
113	5.91965734093497\\
114	4.73595799774703\\
115	3.40093457022108\\
116	1.87328345358337\\
117	0.0891122852454685\\
118	-2.05818917577815\\
119	-4.7671117500475\\
120	-8.47767835406172\\
121	-14.5520450992274\\
122	-36.2768327046756\\
123	-16.0140735778568\\
124	-10.1130773755364\\
125	-6.8603222143193\\
126	-4.68035561182894\\
127	-3.08997212541487\\
128	-1.8761256133494\\
129	-0.925852662209315\\
130	-0.171988428197653\\
131	0.428671747513897\\
132	0.905568529792715\\
133	1.27966722080083\\
134	1.56642567630282\\
135	1.77756525022286\\
136	1.92218079850818\\
137	2.0074650316106\\
138	2.03919733176462\\
139	2.02208306585027\\
140	1.95999480663011\\
141	1.8561473044754\\
142	1.71322654553162\\
143	1.53348623172558\\
144	1.3188206272103\\
145	1.07081988491408\\
146	0.790812094084257\\
147	0.479895019305559\\
148	0.138959616874561\\
149	-0.231293220007681\\
150	-0.630341707594461\\
151	-1.05783739784148\\
152	-1.51360427845582\\
153	-1.99764271545093\\
154	-2.51013847783977\\
155	-3.05147737747776\\
156	-3.62226640261808\\
157	-4.22336265929513\\
158	-4.85591201272211\\
159	-5.5214001187239\\
160	-6.22171967117291\\
161	-6.95925935267964\\
162	-7.7370224671763\\
163	-8.55878706498171\\
164	-9.42932542452179\\
165	-10.3547106019172\\
166	-11.342754309697\\
167	-12.4036492554389\\
168	-13.5509416374135\\
169	-14.8030601346747\\
170	-16.1858322566281\\
171	-17.7368657243156\\
172	-19.5137401131564\\
173	-21.6108137009072\\
174	-24.1983667842356\\
175	-27.6321870271244\\
176	-32.8632470362849\\
177	-43.1855103382077\\
178	-36.9662424680596\\
179	-30.5811269509673\\
180	-27.0313912775554\\
};
\addplot [color=mycolor4, dashed, line width=1.0pt, forget plot]
  table[row sep=crcr]{%
0	-18.2958861993841\\
1	-19.3358386009677\\
2	-20.456221036471\\
3	-21.6778013087258\\
4	-23.0283541219057\\
5	-24.5447524202631\\
6	-26.2734135746241\\
7	-28.2597776850412\\
8	-30.4849823038346\\
9	-32.6026241725561\\
10	-33.4714352602538\\
11	-32.2663748005282\\
12	-30.070766593957\\
13	-27.8796337146862\\
14	-25.9449125608537\\
15	-24.2640127203721\\
16	-22.7898420056383\\
17	-21.4780599381325\\
18	-20.2940913835977\\
19	-19.2120549044696\\
20	-18.2126517049564\\
21	-17.2813872121374\\
22	-16.4072651667839\\
23	-15.581860395468\\
24	-14.7986619125985\\
25	-14.0526031860889\\
26	-13.3397216532804\\
27	-12.6569082611401\\
28	-12.0017205441438\\
29	-11.372241225041\\
30	-10.7669699386913\\
31	-10.1847394231981\\
32	-9.62465004518204\\
33	-9.08601824711118\\
34	-8.56833569480735\\
35	-8.07123673793669\\
36	-7.59447238999486\\
37	-7.13788946259759\\
38	-6.7014138021687\\
39	-6.2850368093479\\
40	-5.88880459575772\\
41	-5.5128092650997\\
42	-5.15718190697634\\
43	-4.82208697013583\\
44	-4.5077177425225\\
45	-4.21429271251343\\
46	-3.94205262183362\\
47	-3.69125804785569\\
48	-3.46218737268137\\
49	-3.25513500947417\\
50	-3.07040976347341\\
51	-2.90833320614106\\
52	-2.7692379358164\\
53	-2.65346558660748\\
54	-2.56136442821881\\
55	-2.49328637182085\\
56	-2.44958315934036\\
57	-2.43060146369802\\
58	-2.43667656315121\\
59	-2.4681241712996\\
60	-2.52522990267065\\
61	-2.60823572969479\\
62	-2.71732263919003\\
63	-2.85258852706643\\
64	-3.01402018654116\\
65	-3.20145806600222\\
66	-3.41455233404391\\
67	-3.6527087564787\\
68	-3.91502307311667\\
69	-4.20020313395513\\
70	-4.50647926967758\\
71	-4.83150557110307\\
72	-5.17225833056314\\
73	-5.52494318475254\\
74	-5.88492949426336\\
75	-6.24673840959512\\
76	-6.60411771068717\\
77	-6.95023779228831\\
78	-7.27803350911577\\
79	-7.58069102744222\\
80	-7.85223730854787\\
81	-8.08814197799442\\
82	-8.28580702502128\\
83	-8.44482129591578\\
84	-8.56690378443913\\
85	-8.65553833355123\\
86	-8.71537976787099\\
87	-8.75155587648164\\
88	-8.76898927028804\\
89	-8.77182942810565\\
90	-8.76304098288353\\
91	-8.74415870735423\\
92	-8.71520097631699\\
93	-8.67472981834912\\
94	-8.62004953500844\\
95	-8.54753841099505\\
96	-8.45310198151831\\
97	-8.33271835919083\\
98	-8.18301924104442\\
99	-8.00182462344188\\
100	-7.78853947655638\\
101	-7.54433798757181\\
102	-7.27210506660034\\
103	-6.97616091291748\\
104	-6.66184074139382\\
105	-6.33502185889596\\
106	-6.00168151575099\\
107	-5.66754116542352\\
108	-5.33782021854755\\
109	-5.01709604047219\\
110	-4.70925134644865\\
111	-4.41748454681655\\
112	-4.14435974974416\\
113	-3.89187762032828\\
114	-3.66155363107315\\
115	-3.45449503727705\\
116	-3.27147163355766\\
117	-3.11297795273268\\
118	-2.97928623762548\\
119	-2.8704904916252\\
120	-2.7865424162236\\
121	-2.72728024633058\\
122	-2.69245152009193\\
123	-2.68173075104492\\
124	-2.69473285810342\\
125	-2.73102308350391\\
126	-2.79012400735092\\
127	-2.87152015830244\\
128	-2.97466062688362\\
129	-3.09896001191164\\
130	-3.24379797128062\\
131	-3.40851760515639\\
132	-3.59242287171464\\
133	-3.79477522237873\\
134	-4.01478964479059\\
135	-4.2516303173964\\
136	-4.50440610948721\\
137	-4.77216620448587\\
138	-5.05389618122328\\
139	-5.34851495570261\\
140	-5.65487306035971\\
141	-5.97175281247656\\
142	-6.29787098838625\\
143	-6.63188466204431\\
144	-6.97240086868883\\
145	-7.31799069769815\\
146	-7.66720828461312\\
147	-8.0186149460089\\
148	-8.37080837730585\\
149	-8.72245642285808\\
150	-9.07233446022025\\
151	-9.41936496915808\\
152	-9.76265745267832\\
153	-10.1015466234843\\
154	-10.4356267398765\\
155	-10.7647802202026\\
156	-11.0891991930625\\
157	-11.4093994106786\\
158	-11.7262268796646\\
159	-12.040858534839\\
160	-12.3547991863874\\
161	-12.6698777265737\\
162	-12.9882461592345\\
163	-13.3123854470283\\
164	-13.6451225565067\\
165	-13.9896635800137\\
166	-14.3496486500105\\
167	-14.729235841417\\
168	-15.1332238206925\\
169	-15.5672273323337\\
170	-16.037926868505\\
171	-16.553426111061\\
172	-17.1237718530763\\
173	-17.7617287670686\\
174	-18.4839715239969\\
175	-19.3129945402514\\
176	-20.2803283707133\\
177	-21.4323069457553\\
178	-22.8412724132503\\
179	-24.6297909758682\\
180	-27.0313912775554\\
};
\end{axis}

\begin{axis}[%
width=3.15in,
height=2.362in,
at={(0in,0in)},
scale only axis,
xmin=0,
xmax=1,
ymin=0,
ymax=1,
axis line style={draw=none},
ticks=none,
axis x line*=bottom,
axis y line*=left
]
\addplot [color=black, dotted, forget plot]
  table[row sep=crcr]{%
0.4	0.51\\
0.51875	0.51\\
};
\draw[<->, color=black] (axis cs:0.517,0.845) -- node[right, align=left, inner sep=0] {18 dB} (axis cs:0.518,0.51);
% \node[below right, align=left, inner sep=0]
% at (rel axis cs:0.525,0.53) {18 dB};
\end{axis}
\end{tikzpicture}%
    \caption{ Components of the full-array pattern of the initial circularly polarized patch array from a full-wave simulation (LHCP \lineleg{mycolor2}, RHCP \lineleg{mycolor4,dashed}).}
    \label{fig:circular_patch_array_before_d}
\end{figure}
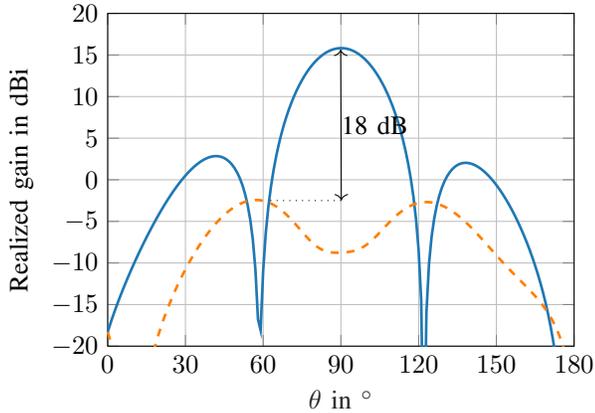

 Now, the feeds are excited with equal amplitude and a full-wave simulation of the initial array is carried out. The LHCP and RHCP components of the directivity in the $\theta$-plane are shown in Fig.~\ref{fig:circular_patch_array_before_d}. A cross-polarization rejection ratio of only $\mathrm{XPR} = \SI{18}{\decibel}$ is observed. This is considered as insufficient for this array and should be improved in the following.
In the conventional approach, this would be done by an optimization process where the edge-lengths of all elements $w_k$ and $l_k$ and their feed positions $p_{\mathrm{W},k}$ and $p_{\mathrm{L},k}$ are varied using full-wave simulations of the entire array. In the following, it is shown how the optimal array can be derived with our proposed approach without running a set of full-wave simulations of the entire array.

\DeclareRobustCommand{\lineleg}[1]{\tikz[baseline=1.7ex,domain=0:0.5] {\node[minimum width=0.5cm, minimum height=0.6em] at (0.25,0.3){};\draw[color=#1,mark repeat=30,mark phase=13,outer sep=1cm] plot (\x,0.3);}}

\subsection{Synthetic Model of the Array Elements}
\label{sec:synth_model}

In order to apply the proposed approach in the following subsection, a model of the circularly polarized patch antenna elements of the array is first derived in this section. Thereby, the synthesis approach from section~\ref{sec:synthesis_approach} is used, where the elements are represented by a model in terms of their modal scattering and radiation behavior.

%To enable a mathematical optimization without performing a high number of full-wave simulations, the elements are represented by the synthetic model from in terms of their modal parameters as degrees of freedom in the following.

%In order to enhance the performance of the array in the coupled case, mutual coupling is taken into account in the following and the elements are modified such that the desired modal configuration is radiated by the elements.

To derive this model, first, it is noted that a circularly polarized patch antenna is well-described already by the first two characteristic modes.
However, since only one port is to be realized, this also means that the elements are non-minimum scattering antennas.

In order to enable the later realization of the synthetic array elements, a synthetic description is developed in which the assignment of the modal parameters to the geometric parameters of the array elements is already taken into account.

On the one hand, the edge-lengths of the patch elements $w_k$ and $l_k$ can be modified.
In the proposed model, this corresponds to the modification of the eigenvalues of the open-circuited antenna $\lambda_1^{\prime{(k)}}$ and $\lambda_2^{\prime{(k)}}$ \cite{Yang2016}.
In the following, the representation as open-circuit scattering matrix $\mathbf{S}_0^\prime$ is therefore used:
\begin{equation}
\label{eq:s0_definition_circular}
    \mathbf{S}_0^{\prime(k)} = \begin{bmatrix}
        s_1^{\prime(k)} & 0 \\
        0 & s_2^{\prime(k)} 
    \end{bmatrix}.
\end{equation}

Furthermore, the positions of the probe feeds $p_{\mathrm{W},k}$ and $p_{\mathrm{L},k}$ can be modified. In the proposed model, this corresponds to the modification of the antenna transmit matrix:
\begin{equation}
\label{eq:t_as_sum_of_t1_and_t2}
    \mathbf{T}^{\prime(k)} = \begin{bmatrix}
    t_1^{\prime(k)} \\
    t_2^{\prime(k)}
    \end{bmatrix}.
\end{equation}
If the antenna is assumed to be matched ($|t_1^{\prime(k)}|^2+|t_2^{\prime(k)}|^2=1$), it means that the probe-feed position mainly influences the ratio $t_2^{\prime(k)}/t_1^{\prime(k)}$. More precisely, since the antenna is also reciprocal, the position of the probe-feed only influences the relative magnitude $|t_2^{\prime(k)}/t_1^{\prime(k)}|$.

This can be seen through the application of the reduction of the degrees of freedom of the antenna elements according to \eqref{eq:condition_formulation2}. By inserting \eqref{eq:t_as_sum_of_t1_and_t2} and \eqref{eq:s0_definition_circular} into \eqref{eq:condition_formulation2}, it is obtained that
\begin{equation}
    s_1^{\prime(k)} t_1^{\prime(k)*}t_2^{\prime(k)}-s_2^{\prime(k)}t_2^{\prime(k)*}t_1^{\prime(k)} = 0
\end{equation}
if the antenna is supposed to be reciprocal. This statement can be reformulated and is actually a statement about the phase between $t_2^{(k)}$ and $t_1^{(k)}$:
\begin{equation}
\label{eq:definition_phase_t1_t2}
    \angle\, \frac{s_2^{\prime(k)}}{s_1^{\prime(k)}} =  -2\, \angle\,  \frac{t_2^{\prime(k)}}{t_1^{\prime(k)}}.
\end{equation}
% 

%Since, in the current example, the phase between $s_2$ and $s_1$ is $\angle (s_2, s_1) = 180^\circ$, the antenna becomes reciprocal for $\angle (t_2, t_1) = \pm 90^\circ$, which is actually the case for a circularly polarized antenna. An example for such an antenna (with $\mathbf{T} = (\mathbf{u}_1+\mathrm{j}\mathbf{u}_2)/\sqrt{2}$) is shown in Fig.~\ref{subfig:scattering_matrix_cm_plate_circular_port}. It is seen that this antenna indeed has a symmetric scattering matrix $\mathbf{S}$ and is therefore reciprocal.

%This is an interesting insight for the synthesis of such antennas.
Therefore, the phase between $t_2^{\prime(k)}$ and $t_1^{\prime(k)}$ is not an independent degree of freedom (up to an uncertainty of 180°) if $s_1^{\prime(k)}$ and $s_2^{\prime(k)}$ are already defined. Translated to geometrical parameters, this means that the port positions $p_{\mathrm{W},k}$ and $p_{\mathrm{L},k}$ only influence the magnitude of the ratio $|t_2^{\prime(k)}/t_1^{\prime(k)}|$, whereby the angle between $t_1^{\prime(k)}$ and $t_2^{\prime(k)}$ is already defined by the edge-lengths of the patches $w_k$ and $l_k$.

The collected insights about the association between the modal and the geometrical parameters will be used later to realize the mathematical synthesized array elements after they have been optimized. For the following mathematical synthesis, the relevant information is that each antenna element has the following three relevant modal degrees of freedom:
\begin{enumerate}
    \item The phase $\angle s_1^{\prime(k)}$.
    \item The phase $\angle s_2^{\prime(k)}$.
    \item The relative magnitude $\left|t_2^{\prime(k)}/t_1^{\prime(k)}\right|$.
\end{enumerate}

\begin{figure}
    \centering
    \begin{subcaptionblock}{.2\textwidth}
        \centering
        %% Creator: Inkscape 1.0.1 (3bc2e813f5, 2020-09-07), www.inkscape.org
%% Accompanies image file 'circular_patch_array_coupling_signal_flow.pdf' (pdf, eps, ps)
\begingroup%
  \makeatletter%
  \providecommand\color[2][]{%
    \errmessage{(Inkscape) Color is used for the text in Inkscape, but the package 'color.sty' is not loaded}%
    \renewcommand\color[2][]{}%
  }%
  \providecommand\transparent[1]{%
    \errmessage{(Inkscape) Transparency is used (non-zero) for the text in Inkscape, but the package 'transparent.sty' is not loaded}%
    \renewcommand\transparent[1]{}%
  }%
  \providecommand\rotatebox[2]{#2}%
  \newcommand*\fsize{\dimexpr\f@size pt\relax}%
  \newcommand*\lineheight[1]{\fontsize{\fsize}{#1\fsize}\selectfont}%
  \ifx\svgwidth\undefined%
    \setlength{\unitlength}{119.05511811bp}%
    \ifx\svgscale\undefined%
      \relax%
    \else%
      \setlength{\unitlength}{\unitlength * \real{\svgscale}}%
    \fi%
  \else%
    \setlength{\unitlength}{\svgwidth}%
  \fi%
  \global\let\svgwidth\undefined%
  \global\let\svgscale\undefined%
  \makeatother%
  \begin{picture}(1,0.85011618)%
    \lineheight{1}%
    \setlength\tabcolsep{0pt}%
    \put(0,0){\includegraphics[width=\unitlength,page=1]{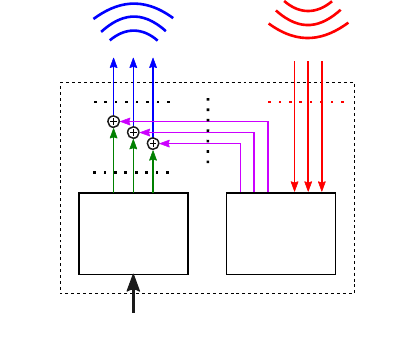}}%
    \put(-0.00531528,0.583242){\makebox(0,0)[lt]{\lineheight{1.25}\smash{\begin{tabular}[t]{l}$\color{myblue}\mathbf{f}^{(k)}$\end{tabular}}}}%
    \put(-0.00513772,0.41318314){\makebox(0,0)[lt]{\lineheight{1.25}\smash{\begin{tabular}[t]{l}$\color{myorange}\mathbf{f}_{\mathrm{T}}^{(k)}$\end{tabular}}}}%
    \put(0.44405832,0.68088056){\makebox(0,0)[lt]{\lineheight{1.25}\smash{\begin{tabular}[t]{l}$\color{myturkis}\mathbf{f}_{\mathrm{sca}}^{(k)}$\end{tabular}}}}%
    \put(0.9037931,0.59087882){\makebox(0,0)[lt]{\lineheight{1.25}\smash{\begin{tabular}[t]{l}$\color{red} \mathbf{a}^{(k)}$\end{tabular}}}}%
    \put(0.60459277,0.25824285){\makebox(0,0)[lt]{\lineheight{1.25}\smash{\begin{tabular}[t]{l}$\mathbf{S}^{\prime(k)}$\end{tabular}}}}%
    \put(0.24314113,0.25832078){\makebox(0,0)[lt]{\lineheight{1.25}\smash{\begin{tabular}[t]{l}$\mathbf{T}^{\prime(k)}$\end{tabular}}}}%
    \put(0.27441474,0.02175342){\makebox(0,0)[lt]{\lineheight{1.25}\smash{\begin{tabular}[t]{l}$v^{(k)}$\end{tabular}}}}%
  \end{picture}%
\endgroup%

        \caption{}\label{subfig:signal_flow}
    \end{subcaptionblock}%
    \qquad\qquad
    \begin{subcaptionblock}{.2\textwidth}
        \centering
        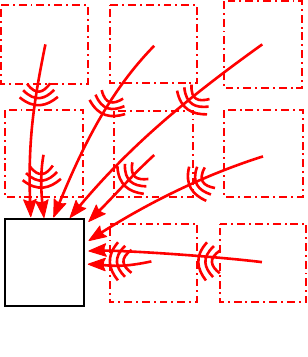
        \caption{}\label{subfig:sources}
    \end{subcaptionblock}%
    \caption{Modal signal flow graph (a) and replacement of  other array elements $l \neq k$ by impressed current sources (b) for the applied coupling model at the $k$\nobreakdash-th array element.}
    \label{fig:circular_patch_array_coupling_signal_flow_and_sources}
\end{figure}

\subsection{Calculation of the Modal Coupling Matrix}

Before the synthesis can take place, the modal coupling matrix $\mathbf{G}$ has to be calculated. Therefore, a patch array without probe feeds using the edge-lengths from the initial array $w_k$ and $l_k$ is set up in the in-house method of moments code \cite{cmc2023}. After calculating the impedance matrix $\mathbf{Z}$ of the entire array and the characteristic modes of the isolated elements, the modal coupling matrix $\mathbf{G}$ is calculated according to \eqref{eq:G_coupling_CM}.

\subsection{Synthesis of the Array}

Now, where the synthetic model for the array elements and the modal coupling matrix are known, the antenna array is optimized to radiate predominantly LHCP in the following. This is done in three steps in the following: first, the optimization goal is formulated mathematically, then an approach to the optimization is given and finally, the results are shown.

\subsubsection{Mathematical Problem Statement}

In the following, two unit vectors 
\begin{equation}
    \mathbf{u}_{\mathrm{L}} = \frac{1}{\sqrt{2}}\begin{bmatrix}
        1\\
        -\mathrm{j}
    \end{bmatrix}
\end{equation}
and
\begin{equation}
    \mathbf{u}_{\mathrm{R}} = \frac{1}{\sqrt{2}}\begin{bmatrix}
        1\\
        \mathrm{j}
    \end{bmatrix}
\end{equation}
are defined, whereby $\mathbf{u}_{\mathrm{L}}$ represents the modal combination on an element that is LHCP and $\mathbf{u}_{\mathrm{R}}$ represents the modal combination on an element that is RHCP.

In order to achieve a modal configuration where the current on all elements is radiating predominantly LHCP, all elements are tuned individually. Representatively, the modal configuration on the $k$\nobreakdash-th array element is tuned in the following.

%\footnote{To obtain a broadside directed beam, all array elements must radiate with the same phase. For simplicity, a zero phase is chosen (without restriction to the generality of the solution).}
Here, the goal is to tune the resulting outgoing field from the $k$\nobreakdash-th array element ${\mathbf{f}^{(k)}}$ such that only LHCP is radiated with a zero phase  and equal amplitude:
\begin{equation}
\label{eq:b_goal}
    \mathbf{f}^{(k)} =\mathbf{u}_{\mathrm{L}}\, q\hspace{0.5cm}\text{for}\hspace{0.5cm}q\in\mathbb{R}^+,
\end{equation}
whereby $q$  is the magnitude of all $\mathbf{f}^{(k)}$.

As shown in Fig.~\ref{subfig:signal_flow}, the resulting outgoing field from the $k$\nobreakdash-th element (represented by ${\mathbf{f}^{(k)}}$) is a sum of the field caused by the excitation of this antenna ${\mathbf{f}_{\mathrm{T}}^{(k)}}$ and the scattered field due to the field incident to the antenna ${\mathbf{f}_{\mathrm{sca}}^{(k)}}$:
\begin{equation}
\label{eq:b_as_sum_of_inc_and_sca}
    {\mathbf{f}^{(k)}} = {\mathbf{f}_{\mathrm{T}}^{(k)}} + {\mathbf{f}_{\mathrm{sca}}^{(k)}} = \mathbf{T}^{\prime(k)} v^{(k)} + \mathbf{S}^{\prime(k)} {\mathbf{a}^{(k)}}.
\end{equation}

For the calculation of the incident field represented by $\mathbf{a}^{(k)}$, the assumption is that every other element $l \neq k$ is already tuned correctly (to only radiate LHCP). This can be imagined by replacing the other antenna elements $l\neq k$ by an imprinted current density that is composed of the desired modal configuration (as indicated by the dashed red lines in Fig.~\ref{subfig:sources}). 
%Mathematically, in terms of the modal weighting coefficients, this means
% \begin{equation}
%     \mathbf{f}^{(l)} = \mathbf{u}_{\mathrm{L}} \, q^{(l)}\hspace{0.5cm}\text{for}\hspace{0.5cm}l\neq k \hspace{0.2cm} \text{and} \hspace{0.2cm} q^{(l)}\in\mathbb{R}^+
% \end{equation}
Mathematically, the weighting coefficients of the field incident to the $k$\nobreakdash-th antenna from the other antennas $l\neq k$ are calculated using the modal coupling matrices $\mathbf{G}^{(k,l)}$ by:
\begin{equation}
\label{eq:example_definition_alpha}
    {\mathbf{a}^{(k)}} = \sum_{l\neq k} \mathbf{G}^{(k,l)} \, \mathbf{f}^{(l)}  = q \sum_{l\neq k} \mathbf{G}^{(k,l)} \, \mathbf{u}_\mathrm{L} = \boldsymbol{\alpha}^{(k)} q.
\end{equation}

In order to achieve the goal defined in \eqref{eq:b_goal}, the antenna transmit vector $\mathbf{T}^{\prime(k)}$, the scattering behavior $\mathbf{S}^{\prime(k)}$ and the excitation $v^{(k)} \in \mathbb{R}^+$ can be adjusted, whereby \eqref{eq:gsm_from_sm} and \eqref{eq:condition_formulation2} have to be fulfilled as side conditions.
%, whereby the incident field represented by $\mathbf{a}^{(k)}$ remains a constant.
%\footnote{Theoretically, it would be possible to allow $v^{(k)} \in \mathbb{C}$ here, but we assume that phase shifts are contained within $\mathbf{T}^{(k)}$ (without restriction to the generality of the solution).}.

\subsubsection{Solution to the Mathematical Problem}

% \begin{figure}
%     \centering
%     \begin{subcaptionblock}{.2\textwidth}
%         \centering
%         \input{images/circular_patch_array_coupling_b_lhcp.pdf_tex}
%         \caption{}\label{subfig:b_lhcp}
%     \end{subcaptionblock}%
%     \qquad\qquad
%     \begin{subcaptionblock}{.2\textwidth}
%         \centering
%         \input{images/circular_patch_array_coupling_b_rhcp.pdf_tex}
%         \caption{}\label{subfig:b_rhcp}
%     \end{subcaptionblock}%
%     \caption{LHCP (a) and RHCP (b) components of the modal weighting coefficients on the $k$\nobreakdash-th array element.}
%     \label{fig:circular_patch_array_coupling_mwc}
% \end{figure}

Since multiple (partially interdependent) parameters ($\mathbf{T}^{\prime(k)}$, $\mathbf{S}^{\prime(k)}$, $v^{(k)}$,  $q${}) can be adjusted to find the optimal solution, an iterative solution procedure is chosen here.  The step-index $\tau$ is introduced therefore and the quantities are renamed to $\mathbf{T}_\tau^{\prime(k)}$, $\mathbf{S}^{\prime(k)}_\tau$, $v^{(k)}_\tau$, $q_\tau$ within this subsection. Especially, it is noted that the scattering behaviour $\mathbf{S}^{\prime(k)}_\tau$ is dependent on the transmit behavior $\mathbf{T}^{\prime(k)}_\tau$ and the open-circuit scattering behavior $\mathbf{S}_{0,\tau}^{\prime(k)}$ according to \eqref{eq:gsm_from_sm}. Here, this fact is considered by iteratively updating $\mathbf{S}^{\prime(k)}_\tau$ in every step according to $\mathbf{T}^{\prime(k)}_{\tau-1}$ and $ \mathbf{S}_{0,\tau-1}^{\prime(k)}$ from the previous step:
\begin{equation}
    \mathbf{S}^{\prime(k)}_\tau = \mathbf{S}_{0,\tau-1}^{\prime(k)}\left(\mathbf{I}- \mathbf{T}^{\prime(k)*}_{\tau-1}\mathbf{T}^\mathrm{\prime(k)T}_{\tau-1}\right).
\end{equation}

Furthermore, to comply with \eqref{eq:condition_formulation2}, $\mathbf{S}_{0,\tau-1}^{\prime(k)}$ is also obtained based on $\mathbf{T}^{\prime(k)}_{\tau-1}$:
\begin{equation}
     \mathbf{S}_{0,\tau-1}^{\prime(k)} = \sigma_k \begin{bmatrix}
        {\mathrm{e}^{2 \cdot \angle t_{1,\tau-1}^{\prime (k)}}} & 0 \\
        0 & {\mathrm{e}^{2 \cdot \angle t_{2,\tau-1}^{\prime (k)}}}
    \end{bmatrix},
\end{equation}
whereby $\sigma_k = \mathrm{j}$ for the vertical elements $k\in \{1, 2, 5, 8, 9\}$ and $\sigma_k = -\mathrm{j}$ for the others.

%\mathbf{S}_0^{\prime (k)}\,\mathbf{T}^{\prime (k)*}\mathbf{T}^{\prime (k)\text{T}} - {\mathbf{T}^{\prime(k)}}\mathbf{T}^{\prime(k){\text{H}}}\mathbf{S}_0^{\prime(k)}\, = \mathbf{0}

% \begin{equation}
%     \mathbf{f}_{\mathrm{sca},\tau}^{(k)} = \mathbf{S}^{\prime(k)}_\tau \boldsymbol{\alpha}^{(k)} q_\tau = \boldsymbol{\zeta}_{\mathrm{sca},\tau}^{(k)} q_\tau
% \end{equation}

% To derive an iteration for $\mathbf{T}^{\prime(k)}_{\tau}$, 

To derive an iteration rule for $\mathbf{T}^{\prime(k)}_{\tau}$, \eqref{eq:b_goal}, \eqref{eq:b_as_sum_of_inc_and_sca} and \eqref{eq:example_definition_alpha} are combined and the following equation is obtained:
\begin{equation}
    q_\tau \, \mathbf{u}_\mathrm{L} = \mathbf{T}^{\prime(k)}_{\tau} v^{(k)}_\tau + \mathbf{S}^{\prime(k)}_\tau \boldsymbol{\alpha}^{(k)} q_\tau.
\end{equation}

% \begin{equation}
%     q_\tau \, \mathbf{u}_\mathrm{L} = \mathbf{T}^{\prime(k)}_{\tau} v^{(k)}_\tau +  \boldsymbol{\zeta}_{\mathrm{sca},\tau}^{(k)} q_\tau
% \end{equation}.

Solving this equation for $\mathbf{T}^{\prime(k)}_{\tau}$, leads to:
\begin{equation}
    \label{eq:T_circ_not_yet_solved}
    \mathbf{T}^{\prime(k)}_{\tau} = \frac{q_\tau}{v^{(k)}_\tau} \, \left(\mathbf{u}_\mathrm{L} - \mathbf{S}^{\prime(k)}_\tau \boldsymbol{\alpha}^{(k)}\right).
\end{equation}

While $q_\tau$ and $v^{(k)}_\tau$ are both not known yet, the fact that $||\mathbf{T}^{\prime(k)}_{\tau}|| = 1$ is used to already obtain the formula for $\mathbf{T}^{\prime(k)}_{\tau}$:
\begin{equation}
    \mathbf{T}^{\prime(k)}_{\tau} = \frac{\mathbf{u}_\mathrm{L} - \mathbf{S}^{\prime(k)}_\tau \boldsymbol{\alpha}^{(k)}}{\left\|\mathbf{u}_\mathrm{L} - \mathbf{S}^{\prime(k)}_\tau \boldsymbol{\alpha}^{(k)}\right\|}.
\end{equation}

Taking the norm $\left\|...\right\|$ on both sides of \eqref{eq:T_circ_not_yet_solved} and bringing $v^{(k)}_\tau$ to the left-hand side of the equation leads to:
\begin{equation}
    \label{eq:v_k_tau}
    v^{(k)}_\tau = q_\tau \, \left\|\mathbf{u}_\mathrm{L} - \mathbf{S}^{\prime(k)}_\tau \boldsymbol{\alpha}^{(k)}\right\|.
\end{equation}

Now, it is assumed that the incident power to all ports is determined by:
\begin{equation}
    \label{eq:v_k_tau_squared_sum_equals_1}
    {\sum\limits_{k = 1}^K {\left| {v_\tau ^{(k)}} \right|} ^2} = 1.
\end{equation}

Combining \eqref{eq:v_k_tau} and \eqref{eq:v_k_tau_squared_sum_equals_1} gives:
\begin{equation}
    {q_\tau } = \frac{1}{{\sqrt {\sum\limits_{k = 1}^K {{{\left\| {{{\bf{u}}_{\rm{L}}} - {\bf{S}}_\tau ^{\prime (k)}{{\boldsymbol{\alpha }}^{(k)}}} \right\|}^2}} } }}.
\end{equation}

Since $q_\tau$ is now known, \eqref{eq:v_k_tau} can be used to calculate also the excitation of all elements $v_\tau ^{(k)}$. Therefore, the iteration rules are now defined for all parameters and the problem can be solved.

% the following three things are used as goals in order of importance:
% \begin{enumerate}
%     \item The RHCP part is cancelled: $ \mathbf{u}_{\mathrm{R}}^\mathrm{H} \mathbf{f}^{(k)}_\tau = 0$ (see Fig.~\ref{subfig:b_rhcp}).
%     \item  To obtain a broadside directed beam, the phase of the LHCP part is set to zero phase:{} $ \operatorname{Im} {\mathbf{u}_{\mathrm{L}}^\mathrm{H} \mathbf{f}^{(k)}_\tau} = 0$ (see Fig.~\ref{subfig:b_lhcp}).\footnote{To obtain a broadside directed beam, all array elements must radiate with the same phase. For simplicity, a zero phase is chosen (without restriction to the generality of the solution).}
%     \item The remaining port power is used to make $\operatorname{Re} {\mathbf{u}_{\mathrm{L}}^\mathrm{H} \mathbf{f}^{(k)}_\tau}$ as large as possible (see Fig.~\ref{subfig:b_lhcp}). 
% \end{enumerate}

% This leads to the formula:
% \begin{equation}
%     \mathbf{T}^{\prime(k)}_{\tau} = - \frac{1}{v^{(k)}_{\tau}} \left(\mathbf{u}_{\mathrm{R}}\mathbf{u}_{\mathrm{R}}^\mathrm{H} \mathbf{f}_{\mathrm{sca},\tau}^{(k)} + \mathrm{j}\mathbf{u}_{\mathrm{L}} \operatorname{Im} \mathbf{u}_{\mathrm{L}}^\mathrm{H} \mathbf{f}_{\mathrm{sca},\tau}^{(k)} \right) + \chi_\tau^{(k)} \mathbf{u}_{\mathrm{L}},
% \end{equation}
% whereby $\chi_\tau^{(k)} \in \mathbb{R}$ is chosen such that $|\mathbf{T}^{\prime(k)}_{\tau}| = 1$.

The transmit vector is initialized as $\mathbf{T}_0^{\prime(k)} = \mathbf{u}_\mathrm{L}$ and the modal configuration of the isolated elements $\mathbf{f}_{\mathrm{T},\tau} = \mathbf{T}_\tau^{\prime(k)} v_\tau ^{(k)}$ is used to formulate a convergence criterion. Convergence is assumed if all $\mathbf{f}_{\mathrm{T},\tau}^{(k)}$ only change by less than one percent compared to $\mathbf{f}_{\mathrm{T},\tau-1}^{(k)}$ in the following sense:
\begin{equation}
    \sum\limits_{k = 1}^K \left\|  \mathbf{f}_{\mathrm{T},\tau}^{(k)} -  \mathbf{f}_{\mathrm{T},\tau-1} ^{(k)} \right\| < 0.01.
\end{equation}
For this example, five steps of iteration are performed until this point is reached.

\subsubsection{Results After Solution}

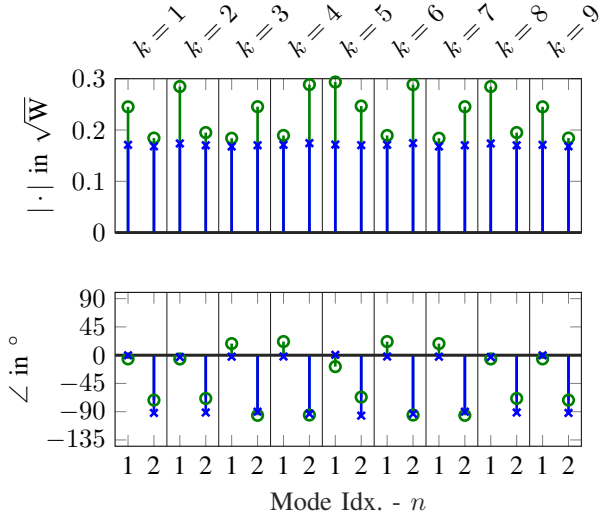
\begin{figure}
    \centering
%\definecolor{mycolor1}{rgb}{0.60000,0.43922,0.67059}%
%\definecolor{mycolor2}{rgb}{0.00392,0.40000,0.36863}%
\definecolor{mycolor1}{rgb}{0, 0.5, 0}%
\definecolor{mycolor2}{rgb}{0,0,1}%
\begin{tikzpicture}

\begin{axis}[%
width=2.441in,
height=0.806in,
at={(0.409in,1.379in)},
scale only axis,
clip=false,
xmin=0.5,
xmax=18.5,
xtick={1,2,3,4,5,6,7,8,9,10,11,12,13,14,15,16,17,18},
xticklabels={{},{},{},{},{},{},{},{},{},{},{},{},{},{},{},{},{},{}},
ylabel={$\left|\,\cdot\,\right|$ in $\sqrt{\text{W}}$},
ymin=0,
ymax=0.3,
axis background/.style={fill=white}
]
\addplot[ycomb, color=mycolor1, line width=1.0pt, mark=o, mark options={solid, mycolor1}, forget plot] table[row sep=crcr] {%
1	0.245231685387074\\
2	0.184334119307953\\
3	0.284758252404982\\
4	0.19501380655247\\
5	0.183978554591999\\
6	0.245498550930678\\
7	0.189385776093072\\
8	0.288531930420964\\
9	0.293534828547722\\
10	0.246771408309129\\
11	0.189374790561966\\
12	0.288539140773717\\
13	0.183989558946758\\
14	0.245490303795081\\
15	0.284751784368966\\
16	0.195023250812598\\
17	0.24523421824\\
18	0.18433074964053\\
};
\addplot[forget plot, color=white!15!black, line width=1.0pt] table[row sep=crcr] {%
0.5	0\\
18.5	0\\
};
\addplot[ycomb, color=mycolor2, line width=1.0pt, mark=x, mark options={solid, mycolor2}, forget plot] table[row sep=crcr] {%
1	0.171131139015318\\
2	0.16823032251119\\
3	0.173792254000497\\
4	0.170223027920996\\
5	0.167910192491308\\
6	0.170276230423643\\
7	0.171161937948873\\
8	0.174445166636115\\
9	0.171660687743299\\
10	0.17050522958451\\
11	0.171158247214364\\
12	0.174447563583503\\
13	0.167917582233235\\
14	0.170283580646637\\
15	0.173792205438428\\
16	0.170227081773675\\
17	0.171130991419851\\
18	0.168230242511657\\
};
\addplot[forget plot, color=white!15!black, line width=1.0pt] table[row sep=crcr] {%
0.5	0\\
18.5	0\\
};
\addplot [color=white!15!black, forget plot]
  table[row sep=crcr]{%
2.5	0\\
2.5	0.3\\
};
\node[right, align=left, inner sep=0, rotate=45]
at (axis cs:1.3,0.35) {$k = 1$};
\addplot [color=white!15!black, forget plot]
  table[row sep=crcr]{%
4.5	0\\
4.5	0.3\\
};
\node[right, align=left, inner sep=0, rotate=45]
at (axis cs:3.3,0.35) {$k = 2$};
\addplot [color=white!15!black, forget plot]
  table[row sep=crcr]{%
6.5	0\\
6.5	0.3\\
};
\node[right, align=left, inner sep=0, rotate=45]
at (axis cs:5.3,0.35) {$k = 3$};
\addplot [color=white!15!black, forget plot]
  table[row sep=crcr]{%
8.5	0\\
8.5	0.3\\
};
\node[right, align=left, inner sep=0, rotate=45]
at (axis cs:7.3,0.35) {$k = 4$};
\addplot [color=white!15!black, forget plot]
  table[row sep=crcr]{%
10.5	0\\
10.5	0.3\\
};
\node[right, align=left, inner sep=0, rotate=45]
at (axis cs:9.3,0.35) {$k = 5$};
\addplot [color=white!15!black, forget plot]
  table[row sep=crcr]{%
12.5	0\\
12.5	0.3\\
};
\node[right, align=left, inner sep=0, rotate=45]
at (axis cs:11.3,0.35) {$k = 6$};
\addplot [color=white!15!black, forget plot]
  table[row sep=crcr]{%
14.5	0\\
14.5	0.3\\
};
\node[right, align=left, inner sep=0, rotate=45]
at (axis cs:13.3,0.35) {$k = 7$};
\addplot [color=white!15!black, forget plot]
  table[row sep=crcr]{%
16.5	0\\
16.5	0.3\\
};
\node[right, align=left, inner sep=0, rotate=45]
at (axis cs:15.3,0.35) {$k = 8$};
\addplot [color=white!15!black, forget plot]
  table[row sep=crcr]{%
18.5	0\\
18.5	0.3\\
};
\node[right, align=left, inner sep=0, rotate=45]
at (axis cs:17.3,0.35) {$k = 9$};
\end{axis}

\begin{axis}[%
width=2.441in,
height=0.806in,
at={(0.409in,0.26in)},
scale only axis,
xmin=0.5,
xmax=18.5,
xtick={1,2,3,4,5,6,7,8,9,10,11,12,13,14,15,16,17,18},
xticklabels={{1},{2},{1},{2},{1},{2},{1},{2},{1},{2},{1},{2},{1},{2},{1},{2},{1},{2}},
xlabel style={font=\color{white!15!black}},
xlabel={Mode Idx. - $n$},
ylabel={$\angle$ in $^\circ$},
ymin=-145,
ymax=100,
ytick={-135,  -90,  -45,    0,   45,   90},
axis background/.style={fill=white}
]
\addplot[ycomb, color=mycolor1, line width=1.0pt, mark=o, mark options={solid, mycolor1}, forget plot] table[row sep=crcr] {%
1	-5.95110656278537\\
2	-71.5633372697493\\
3	-6.02485453291148\\
4	-68.908275336409\\
5	18.4122749437278\\
6	-95.8326816539665\\
7	21.6350543876198\\
8	-95.3665835452171\\
9	-18.2265726573379\\
10	-66.7821030320907\\
11	21.6355595805276\\
12	-95.3645522271454\\
13	18.4095611427748\\
14	-95.8326940366951\\
15	-6.02575400141135\\
16	-68.9094865046394\\
17	-5.95054121863469\\
18	-71.5630364841065\\
};
\addplot[forget plot, color=white!15!black, line width=1.0pt] table[row sep=crcr] {%
0.5	0\\
18.5	0\\
};
\addplot[ycomb, color=mycolor2, line width=1.0pt, mark=x, mark options={solid, mycolor2}, forget plot] table[row sep=crcr] {%
1	-0.240455102115583\\
2	-91.9692863156639\\
3	-2.90236450638064\\
4	-91.2826293268678\\
5	-2.37772455405874\\
6	-89.993348611211\\
7	-2.12326069796832\\
8	-93.3854587612207\\
9	0.556132830102918\\
10	-96.2897617123388\\
11	-2.12439342038142\\
12	-93.3860836329209\\
13	-2.37609385612577\\
14	-89.992508288783\\
15	-2.90193955371691\\
16	-91.2816989052073\\
17	-0.240008331878296\\
18	-91.9701908313866\\
};
\addplot[forget plot, color=white!15!black, line width=1.0pt] table[row sep=crcr] {%
0.5	0\\
18.5	0\\
};
\addplot [color=white!15!black, forget plot]
  table[row sep=crcr]{%
2.5	-145\\
2.5	100\\
};
\addplot [color=white!15!black, forget plot]
  table[row sep=crcr]{%
4.5	-145\\
4.5	100\\
};
\addplot [color=white!15!black, forget plot]
  table[row sep=crcr]{%
6.5	-145\\
6.5	100\\
};
\addplot [color=white!15!black, forget plot]
  table[row sep=crcr]{%
8.5	-145\\
8.5	100\\
};
\addplot [color=white!15!black, forget plot]
  table[row sep=crcr]{%
10.5	-145\\
10.5	100\\
};
\addplot [color=white!15!black, forget plot]
  table[row sep=crcr]{%
12.5	-145\\
12.5	100\\
};
\addplot [color=white!15!black, forget plot]
  table[row sep=crcr]{%
14.5	-145\\
14.5	100\\
};
\addplot [color=white!15!black, forget plot]
  table[row sep=crcr]{%
16.5	-145\\
16.5	100\\
};
\addplot [color=white!15!black, forget plot]
  table[row sep=crcr]{%
18.5	-145\\
18.5	100\\
};
\addplot [color=white!15!black, forget plot]
  table[row sep=crcr]{%
2.5	0\\
2.5	0.3\\
};
\addplot [color=white!15!black, forget plot]
  table[row sep=crcr]{%
4.5	0\\
4.5	0.3\\
};
\addplot [color=white!15!black, forget plot]
  table[row sep=crcr]{%
6.5	0\\
6.5	0.3\\
};
\addplot [color=white!15!black, forget plot]
  table[row sep=crcr]{%
8.5	0\\
8.5	0.3\\
};
\addplot [color=white!15!black, forget plot]
  table[row sep=crcr]{%
10.5	0\\
10.5	0.3\\
};
\addplot [color=white!15!black, forget plot]
  table[row sep=crcr]{%
12.5	0\\
12.5	0.3\\
};
\addplot [color=white!15!black, forget plot]
  table[row sep=crcr]{%
14.5	0\\
14.5	0.3\\
};
\addplot [color=white!15!black, forget plot]
  table[row sep=crcr]{%
16.5	0\\
16.5	0.3\\
};
\addplot [color=white!15!black, forget plot]
  table[row sep=crcr]{%
18.5	0\\
18.5	0.3\\
};
\end{axis}
\end{tikzpicture}%
    \caption{ Magnitude $|\cdot|$ in $\sqrt{\text{W}}$ (top) and angle $\angle$ in $^\circ$  (bottom) of the modal weighting coefficients for the $n$\nobreakdash-th mode on the $k$\nobreakdash-th element for the isolated case  $f_{\mathrm{T},n}^{(k)}$~(\lineleg{mycolor1,mark=o,line width=1}) and the coupled case $f_{n}^{(k)}$~(\lineleg{mycolor2,mark=x,line width=1}) in the model of the final circularly polarized patch array.}
    \label{fig:circular_patch_array_model}
\end{figure}

The resulting modal weighting coefficients are shown in Fig.~\ref{fig:circular_patch_array_model}. It is seen that while the active modal configuration $\mathbf{f}^{(k)}$ now has equal magnitude and the modes are $\SI{90}{\degree}$ out-of-phase, the modal configuration of the isolated elements $\mathbf{f}_\mathrm{T}^{(k)}$ is no longer just LHCP. It contains both LHCP and RHCP parts. We call this `modal pre-distortion'.

In table~\ref{tab:circ_patch_geometry_after}, the final angles $\angle s_1^{\prime(k)}$ and $\angle s_2^{\prime(k)}$, the final magnitudes of the transmit matrix entries $|t_1^{\prime(k)}|$ and $|t_2^{\prime(k)}|$ and the incident wave port amplitudes $v^{(k)}$ are given.

\setlength{\tabcolsep}{0.4em}
\renewcommand{\arraystretch}{1.2}
\newcommand\xrowht[2][0]{\addstackgap[.5\dimexpr#2\relax]{\vphantom{#1}}}
\begin{table}
\caption{Parameters of the Improved Circularly Polarized Patch Array}
\label{tab:circ_patch_geometry_after}
\begin{tabular}{|c*{9}{|>{\centering\arraybackslash}p{0.55cm}}|}
\hline
\rule{0pt}{15pt}
$k$
\rule{0pt}{15pt} & 1    & 2    & 3    & 4    & 5    & 6    & 7    & 8    & 9     \\[5pt] \hline \hline

\multicolumn{10}{|c|}{Modal Parameters of the Elements}      \\ \hline\hline
$\begin{tabular}[c]{@{}c@{}}$ \angle s_1^{\prime (k)}$\\ in ${}^\circ$\end{tabular} $ & 80 & 84 & -50 & -43 & 56 & -43 & -50 & 84 & 80   \\ \hline
$\begin{tabular}[c]{@{}c@{}}$ \angle s_2^{\prime (k)}$\\ in ${}^\circ$\end{tabular} $        & -50 & -45 & 81 & 86 & -35 & 86 & 81 & -45 & -50  \\ \hline\xrowht{15pt}
$\left|t_1^{\prime (k)}\right|$                           & 0.80 & 0.83 & 0.60 & 0.54 & 0.78 & 0.54 & 0.60 & 0.83 & 0.80 \\ \hline\xrowht{15pt}
$\left|t_2^{\prime (k)}\right|$                           & 0.60 & 0.55 & 0.80 & 0.84 & 0.62 & 0.84 & 0.80 & 0.55 & 0.60 \\ \hline \hline
\multicolumn{10}{|c|}{Excitation of the Elements}      \\ \hline\hline\xrowht{15pt}
$v^{(k)}$ & 0.31 & 0.34 & 0.31 & 0.34 & 0.38 & 0.34 & 0.31 & 0.34 & 0.31 \\ \hline  \hline
\multicolumn{10}{|c|}{Parameters for the Geometric Realization of the Elements}      \\ \hline\hline\xrowht{15pt}
\begin{tabular}[c]{@{}c@{}}$w_k$\\ in mm\end{tabular}           & 4.15 & 4.1  & 4.15 & 4.1  & 4.05 & 4.1  & 4.15 & 4.1  & 4.15 \\ \hline
\begin{tabular}[c]{@{}c@{}}$l_k$\\ in mm\end{tabular}           & 4.8  & 4.8  & 4.8  & 4.8  & 5    & 4.8  & 4.8  & 4.8  & 4.8  \\ \hline
\begin{tabular}[c]{@{}c@{}}$p_{\text{W},k}$\\ in \si{\micro\meter}\end{tabular} & 950  & 950  & 950  & 950  & 1150 & 950  & 950  & 950  & 950  \\ \hline
\begin{tabular}[c]{@{}c@{}}$p_{\text{L},k}$\\ in \si{\micro\meter}\end{tabular} & 1100 & 1100 & 1100 & 1100 & 1550 & 1100 & 1100 & 1100 & 1100 \\ \hline
\end{tabular}
\end{table}

\subsection{Geometric Realization of the Elements}

In order to realize the synthesized elements in the array using actual patch antennas with probe feeds,{} each element is individually tuned (without presence of other elements) to achieve the desired modal configuration (from the isolated case). The edge lengths of the patches ($w_k$ and $l_k$) are thereby tuned  first such that $\angle s_1^{\prime(k)}$ and $\angle s_2^{\prime(k)}$ are achieved.{} Afterwards, the probe feed positions ($p_{\text{W}k}$ and $p_{\text{L}k}$) are adjusted such that the transmit matrix entries $t_1^{\prime(k)}$ and $t_2^{\prime(k)}$ are achieved. The resulting geometrical parameters are also shown in Table~\ref{tab:circ_patch_geometry_after}.

\begin{figure}
    \centering
    \input{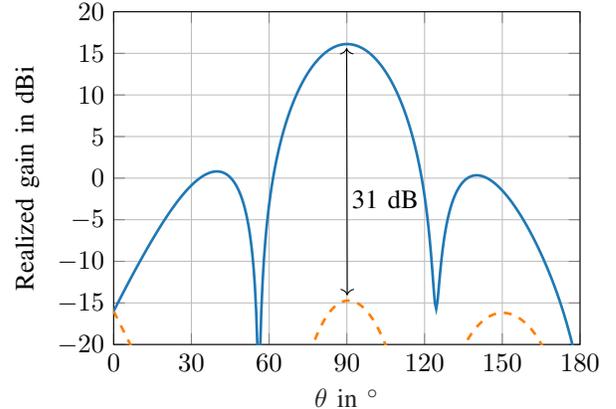}
    \caption{ Components of the full-array pattern of the final circularly polarized patch array from a full-wave simulation (LHCP \lineleg{mycolor2}, RHCP \lineleg{mycolor4,dashed}).}
    \label{fig:circular_patch_array_after_d}
\end{figure}

After the elements have been adjusted to radiate the desired modal configuration isolated from each other, a full-wave simulation of the entire array is conducted. The full-array pattern is shown in Fig.~\ref{fig:circular_patch_array_after_d}. It is seen that the RHCP is now suppressed by $\mathrm{XPR} = \SI{31}{\decibel}$ in comparison to the LHCP. This is considered to be a significant improvement over the cross-polarization rejection ratio of $\SI{18}{\decibel}$ from the initial array.

To summarize, the cross-polarization rejection of a circularly polarized patch array is improved by modification of the individual array elements. This is achieved with the help of a modal coupling model based on the characteristic modes of the individual array elements. The mathematical model is used to calculate a pre-distortion of the isolated modal configuration that leads to the desired modal configuration in the active case (when all array elements are excited).

Finally, it is noted that a systematic design procedure was shown, in which the parameters of the array elements to be controlled, such as edge lengths and port position, were known from the outset. However, this approach can also be applied to other elements where the tuning possibilities of the parameters are not immediately obvious. Other publications have already shown that characteristic modes can be used to demonstrate clear and meaningful relationships between the modal parameters and the corresponding geometrical parameters \cite{Li2022}. These relationships, which were previously only helpful in the design of individual antennas, can now also be used effectively in the development of complex antenna arrays as a result of the approach proposed in this work.

\annotateChangesStart
\section{Discussion}

After this example, it makes sense to return to the approach in general to discuss the two fundamental assumptions of the approach again. The first assumption is that a small number of modes is sufficient to generate a sufficiently accurate coupling model. The second assumption is that the modal coupling matrix does not change significantly due to the geometry changes required for the synthesis. This obviously leads to the questions of how many modes should be considered on the elements and how much geometry change is considered acceptable for the model to remain valid. Since the potential field of application for the proposed approach is large, it is not possible to find generic answers to these questions. Instead, it is proposed to answer these questions for the specific problem in question.

For the presented example of the circularly polarized patch array, the configuration predicted by the model also shows a significant improvement in a full-wave simulation, which means that two modes per element were sufficient and the change in geometry was not relevant for the model to become inaccurate. However, it should be mentioned that other examples may require to include more significant modes while there is no indication that it must be unpractically many modes. Furthermore, the model can become too inaccurate for the requirements of some other examinations due to the changes in geometry. In these cases, the modal coupling matrix can be recalculated iteratively.

\annotateChangesEnd

\section{Conclusion}
\label{sec:conclusion}

An approach to the synthesis of antenna arrays using synthetic array elements that can be controlled mathematically is proposed. Using this approach, the coupling between the elements in the synthetic antenna array is taken into account and an optimized array is designed.

First, the generalized scattering matrix in the basis of characteristic modes of the antenna is introduced as a mathematical model for the array elements.
Here, a relationship is found that allows to formulate the generalized scattering matrix of the array elements in terms of the eigenvalues of the element geometry and the modal radiation behavior of the antenna.
This model is particularly useful as it allows a separate understanding of the influence of the antenna geometry and the radiation behaviour of the antenna.

Based on this model, a synthesis approach is proposed where modal parameters (like the eigenvalues and the antenna transmit matrix) are selected as degrees of freedom and controlled mathematically without the need for an electromagnetic implementation, impedance matching and so on.
Finally, the mathematically synthesized elements are implemented with actual antenna elements that resemble the behavior of the mathematically synthesized elements.

As an example, the modal configuration of a circularly polarized patch array is optimized such that the cross-polarization rejection of the array is significantly improved without performing multiple full-wave simulations of the array. A final full-wave simulation shows that the synthetically calculated modal configuration is actually achievable and that the cross polarization rejection of the array is indeed improved.

In further investigations, the proposed synthesis approach can be applied to more real-world design problems, such as e.g. the decoupling of multi-mode multi-port antennas in an array \cite{Manteuffel2016,Peitzmeier2018,Peitzmeier2019a,Moerlein2021,Manteuffel2022,Moerlein2022,Peitzmeier2022} or the design of electronically steerable parasitic radiators \cite{Harrington1978,Schlub2003}.

% or


\appendices

\section{Proof of the Losslessness of (\ref*{eq:gsm_from_sm})}
\label{appendix:scatterer_to_antenna}

A lossless and reciprocal scatterer (without any ports) is described by the scattering matrix ${{\mathbf{S}}_0}$. Since the scatterer is lossless, ${{\mathbf{S}}_0}$ is unitary
\begin{equation}
\label{eq:losslessness_scatterer}
    {\mathbf{I}} = {\mathbf{S}}_0^{\text{H}}{{\mathbf{S}}_0},
\end{equation}
and since the scatterer is reciprocal, ${{\mathbf{S}}_0}$ is symmetric
\begin{equation}
    {{\mathbf{S}}_0} = {\mathbf{S}}_0^{\text{T}}.
\end{equation}

Let ${\mathbf{\Psi}}$ be a generalized scattering matrix of a matched and reciprocal antenna with uncoupled ports defined by:
\begin{equation}
    {\mathbf{\Psi}} = \begin{bmatrix}
  {{\mathbf{S}} }& \mathbf{T} \\ 
  {\mathbf{R} }& {\mathbf{\Gamma }}
\end{bmatrix} = \begin{bmatrix}
  \mathbf{S} & \mathbf{T} \\ 
  \mathbf{T}^{\text{T}} &\mathbf{0}
\end{bmatrix},
\end{equation}
whereby the scattering properties of the antenna $\mathbf{S}$ are similar to the aforesaid scatterer (and its scattering matrix $\mathbf{S}_0$) except that ports (represented by $\mathbf{T}$) have been introduced using:
\begin{equation}
\label{eq:definition_antenna_scattering}
    \mathbf{S} = \mathbf{S}_0 - \mathbf{S}_0\,{\mathbf{T}}^{*}{\mathbf{T}}^{\mathrm{T}}.
\end{equation}

In order for the antenna to be lossless, the matrix $\mathbf{\Psi}$ must be unitary:
\begin{equation}
{\mathbf{\Psi}^{\text{H}}}\mathbf{\Psi} = \begin{bmatrix}
  {{{\mathbf{S}}^{\text{H}}}{\mathbf{S}} + {{\mathbf{T}}^*}{{\mathbf{T}}^{\text{T}}}}&{{{\mathbf{S}}^{\text{H}}}{\mathbf{T}}} \\ 
  {{{\mathbf{T}}^{\text{H}}}{\mathbf{S}}}&{{{\mathbf{T}}^{\text{H}}}{\mathbf{T}}} 
\end{bmatrix}\mathop {=}\limits^! {\mathbf{I}}.
\end{equation}

This is shown block-wise. Since the antenna ports are assumed to be matched and uncoupled,
\begin{equation}
\label{eq:tm_is_matched_and_uncoupled}
    {{{\mathbf{T}}^{\text{H}}}{\mathbf{T}}} = \mathbf{I},
\end{equation}
so that the bottom-right block is fulfilled by definition.

The top-left block
\begin{equation}
    {\mathbf{I}} \mathop = \limits^! {{{\mathbf{S}}^{\text{H}}}{\mathbf{S}} + {{\mathbf{T}}^*}{{\mathbf{T}}^{\text{T}}}}
\end{equation}
is transformed using \eqref{eq:definition_antenna_scattering} into
\begin{equation}
    {\mathbf{I}}\mathop  = \limits^! {\left( {{{\mathbf{S}_0}} - {{\mathbf{S}_0}}\,{\mathbf{T}}^{ *}{\mathbf{T}}^{ {\text{T}}}} \right)^{\text{H}}}\left( {{{\mathbf{S}_0}} - {{\mathbf{S}_0}}\,{\mathbf{T}}^{ *}{\mathbf{T}}^{ {\text{T}}}} \right) + {\mathbf{T}}^{ {\text{*}}}{\mathbf{T}}^{ {\text{T}}}.
\end{equation}
Using \eqref{eq:losslessness_scatterer}, the equation
\begin{equation}
    {\mathbf{I}}\mathop  = \limits^! {\mathbf{I}} - {\mathbf{T}}^{ {\text{*}}}{\mathbf{T}}^{ {\text{T}}} + {\mathbf{T}}^{ {\text{*}}}{\mathbf{T}}^{ {\text{T}}}{\mathbf{T}}^{ {\text{*}}}{\mathbf{T}}^{ {\text{T}}}
\end{equation}
is obtained. Using \eqref{eq:tm_is_matched_and_uncoupled}, the equation can be rewritten as:
\begin{equation}
    {\mathbf{I}}\mathop  = \limits^! {\mathbf{I}} - {\mathbf{T}}^{ {\text{*}}}{\mathbf{T}}^{ {\text{T}}} + \,{\mathbf{T}}^{ {\text{*}}}{\mathbf{T}}^{ {\text{T}}} = \mathbf{I}. \hspace{0.5cm}\qedsymbol
\end{equation}
For the off-diagonal blocks, it is necessary to show:
\begin{equation}
    \mathbf{0}\mathop  = \limits^! {\mathbf{S}}^{{\text{H}}}{\mathbf{T}}.
\end{equation}
Since ${\mathbf{S}}^{{\text{T}}} = {\mathbf{S}}$ is assumed, this is equal to:
\begin{equation}
    \mathbf{0}\mathop  = \limits^! {\mathbf{S}}^{{\text{*}}}{\mathbf{T}}.
\end{equation}
Using \eqref{eq:definition_antenna_scattering}, this becomes:
\begin{equation}
{\mathbf{0}}\mathop  = \limits^! {\mathbf{S}}_0 ^{\text{*}}{\mathbf{T}} - {\mathbf{S}}_0 ^{\text{*}}{\mathbf{T}}{{\mathbf{T}}^{\text{H}}}{\mathbf{T}}
\end{equation}
Using \eqref{eq:tm_is_matched_and_uncoupled}, this becomes:
\begin{equation}
    {\mathbf{0}}\mathop  = \limits^! {\mathbf{S}}_0^{\text{*}}{\mathbf{T}} - {\mathbf{S}}_0^{\text{*}}{\mathbf{T}} = {\mathbf{0}}. \hspace{0.5cm}\qedsymbol
\end{equation}

\section{Modal~Coupling~Matrix~for~Characteristic~Modes in Terms of The Impedance Operator Z}
\label{sec:G_calculation}

The boundary condition at the surface of a perfectly conducting scattering object is described by a sum of tangential components of the externally incident field $\mathbf{E}_{\mathrm{inc,ext}}$ and the scattered field $\mathbf{E}_{\mathrm{S}}$:
\begin{equation}
    \left[\mathbf{E}_{\mathrm{inc,ext}} + \mathbf{E}_{\mathrm{S}}\right]_{\mathrm{tan}} = \mathbf{0},
\end{equation}
whereby the scattered field is defined by $\mathbf{E}_\mathrm{S} = - Z \mathbf{J}$ using the impedance operator $Z$ and the surface current distribution $\mathbf{J}$ \cite{Harrington1971}.

Now, in the array case, the surface current distribution $\mathbf{J}$ is composed by the surface current distributions of all array elements $\mathbf{J}^{(l)}$:
\begin{equation}
    \left[\mathbf{E}_{\mathrm{inc,ext}} - \sum_l Z \mathbf{J}^{(l)} \right]_{\mathrm{tan}} = \mathbf{0}
\end{equation}

This equation can be rearranged, such that the scattering at the $k$-th element is investigated and the other $Z \mathbf{J}^{(l)}$ for $l\neq k$ are moved into $\widetilde{\mathbf{E}}_{\mathrm{inc}}^{(k)}$, which is the incident field to the $k$-th element:
\begin{equation}
    \left[\widetilde{\mathbf{E}}_{\mathrm{inc}}^{(k)} - Z \mathbf{J}^{(k)} \right]_{\mathrm{tan},k} = \mathbf{0},
\end{equation}
whereby
\begin{equation}
    \widetilde{\mathbf{E}}_{\mathrm{inc}}^{(k)} = \mathbf{E}_{\mathrm{inc,ext}} - \sum_{l \neq k} Z \mathbf{J}^{(l)}.
\end{equation}

The current on every element $\mathbf{J}^{(l)}$ is expanded into the characteristic modes of the $l$\nobreakdash-th element using:
\begin{equation}
    \mathbf{J}^{(l)} = \sum_{n^\prime} \mathbf{J}_{n^\prime}^{(l)} f_{n^\prime}^{(l)}.
\end{equation}

Finally, using \cite[eq. (9)]{Grundmann2021}, the modal coupling coefficient $G_{n,n^\prime}^{(k,l)}$ can be defined as a relation between the modal weighting coefficient $f_{n^\prime}^{(l)}$ of the $n^\prime$\nobreakdash-th outgoing mode at the $l$\nobreakdash-th array element and the modal weighting coefficient $a_n^{(k)}$ of the $n$\nobreakdash-th incoming mode at the $k$\nobreakdash-th element:
\begin{align}
\begin{split}
    a_n^{(k)} &= - \frac{1}{2} \left<\mathbf{J}_n^{(k)}, \widetilde{\mathbf{E}}_{\mathrm{inc}}^{(k)}\right> \\
    &= - \frac{1}{2} \left<\mathbf{J}_n^{(k)},\mathbf{E}_{\mathrm{inc,ext}}\right> +  \sum_{l \neq k} \sum_{n^\prime} \underbrace{\frac{1}{2} \left<\mathbf{J}_n^{(k)}, Z\mathbf{J}_{n^\prime}^{(l)}\right>}_{G_{n,n^\prime}^{(k,l)}} f_{n^\prime}^{(l)}.
\end{split}
\end{align}

\ifCLASSOPTIONcaptionsoff
  \newpage
\fi



\bibliographystyle{IEEEtran}
\bibliography{IEEEabrv,lit}

\begin{IEEEbiography}[{\includegraphics[width=1in,height=1.25in,clip,keepaspectratio]{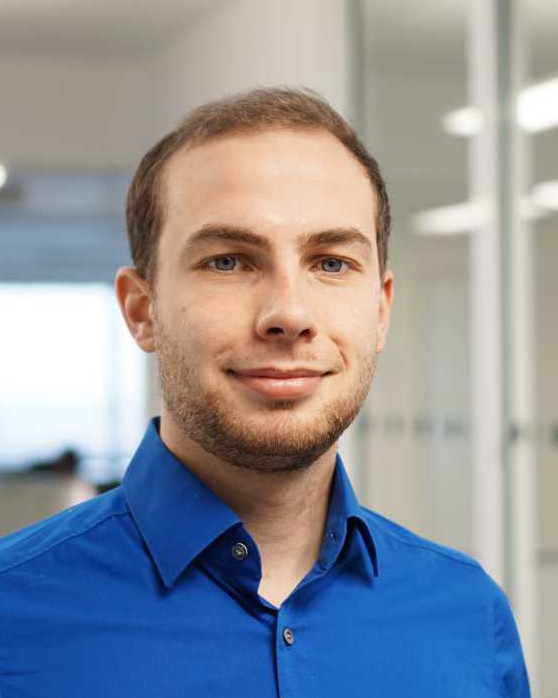}}]{Leonardo Mörlein}
Leonardo Mörlein (Graduate Student Member, IEEE) was born in 1994 in Würzburg, Germany. He received the B.Sc. and M.Sc. degrees in electrical engineering from Leibniz University Hannover, Hannover, Germany, in 2017 and 2020, respectively. He is currently a Research Assistant with the Institute of Microwave and Wireless Systems, Leibniz University Hannover. His current research focuses on the use of multi-port multi-mode antennas in beamforming scenarios. Further research interests include antenna integration, the use of modal decompositions and channel modeling.
\end{IEEEbiography}

\begin{IEEEbiography}[{\includegraphics[width=1in,height=1.25in,clip,keepaspectratio]{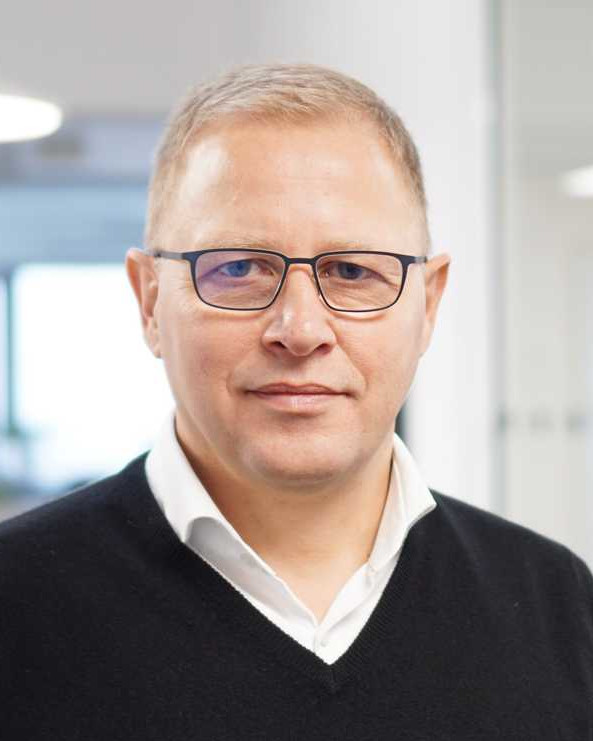}}]{Dirk Manteuffel} (Member, IEEE) was born in Issum, Germany, in 1970. He received the Dipl.-Ing. and Dr.-Ing. degrees in electrical engineering from the University of Duisburg–Essen, Duisburg, Germany, in 1998 and 2002, respectively.

From 1998 to 2009, he was with IMST, Kamp-Lintfort, Germany. As a Project Manager, he was responsible for industrial antenna development and advanced projects in the field of antennas and electromagnetic (EM) modeling. From 2009 to 2016, he was a Full Professor of wireless communications at Christian-Albrechts-University, Kiel, Germany. Since June 2016, he has been a Full Professor and the Executive Director of the Institute of Microwave and Wireless Systems, Leibniz University Hannover, Hannover, Germany. His research interests include electromagnetics, antenna integration and EM modeling for mobile communications and biomedical applications.

Dr. Manteuffel was a director of the European Association on Antennas and Propagation from 2012 to 2015. He served on the Administrative Committee (AdCom) of IEEE Antennas and Propagation Society from 2013 to 2015 and as an Associate Editor of the IEEE Transactions on Antennas and Propagation from 2014 to 2022. Since 2009 he has been an appointed member of the committee "Antennas" of the German VDI-ITG.
\end{IEEEbiography}

%\newpage


%\vfill


\end{document}